\documentclass[times,review,preprint,authoryear,10.5pt]{elsarticle}

\usepackage{jasr}
\usepackage{framed,multirow}

\usepackage{amssymb,bm}
\usepackage{latexsym}
\usepackage{mathtools, cuted}
\usepackage[switch]{lineno}

\usepackage{url}
\usepackage{xcolor}
\definecolor{newcolor}{rgb}{.8,.349,.1}

\usepackage[ruled,vlined]{algorithm2e}

\usepackage{caption}
\usepackage{subcaption}

\journal{Advances in Space Research}

\begin{document}

\verso{Francesca Scala \textit{et al}}

\begin{frontmatter}

\title{Design of optimal low-thrust manoeuvres for remote sensing multi-satellite formation flying in low Earth orbit\tnoteref{tnote1}}%

\author[1]{Francesca \snm{Scala}\corref{cor1}}
\cortext[cor1]{Corresponding author}
\ead{francesca1.scala@polimi.it}
\author[1]{Gabriella \snm{Gaias}}
\ead{gabriella.gaias@polimi.it}
\author[1]{Camilla \snm{Colombo}}
\ead{camilla.colombo@polimi.it}
\author[2]{Manuel \snm{Martin-Neira}}
\ead{manuel.martin-neira@esa.int}

\address[1]{Politecnico di Milano, via La Masa 34, 20156, Milan, Italy}
\address[2]{ESA-ESTEC, Keplerlaan 1, 2200 AG Noordwijk, The Netherlands}


\begin{abstract}
This paper presents a strategy for optimal manoeuvre design of multi-satellite formation flying in low Earth orbit environment, with the aim of providing a tool for mission operation design. The proposed methodology for formation flying manoeuvres foresees a continuous low-thrust control profile, to enable the operational phases. The design is performed starting from the dynamic representation described in the relative orbital elements, including the main orbital perturbations effects. It also exploits an interface with the classical radial-transversal-normal description to include the maximum delta-v limitation and the safety condition requirements. The methodology is applied to a remote sensing mission study, Formation Flying L-band Aperture Synthesis, for land and ocean application, such as a potential high-resolution Soil Moisture and Ocean Salinity (SMOS) follow-on mission, as part of a European Space Agency mission concept study. Moreover, the results are applicable to a wide range of low Earth orbit missions, exploiting a distributed system, and in particular to Formation Flying L-band Aperture Synthesis (FFLAS) as a follow-on concept to SMOS. 
\end{abstract}


\end{frontmatter}



\section{Introduction}
\label{sec:Introduction}

Distributed systems have increased their importance in space application in the last decade. The idea of exploiting the benefits of distributed spacecraft for a common objective can improve the mission performances, and exploiting formation flying for Earth observation could provide an unprecedented advantage to the field in terms of higher spatial resolution \citep{leitner2004formation,bandyopadhyay2016review}. Remote sensing for Earth applications is typically applied in Low Earth Orbit (LEO), resulting in a small field-of-view or limited angular resolution of the scientific instrument onboard a single-satellite mission. Formation flying gives the possibility to increase the payload performances in terms of spatial and temporal resolution and field of view as it increases the effective size of the instrument \citep{krieger2007tandem,moreira2015tandem}. During the operational life of a mission, different phases may require different geometries of the satellites in the formation. Consequently, the design of the mission requires a clear definition of reconfiguration strategies for multiple satellites, for the definition of the performances. 
In particular, this paper presents an algorithm for a fuel-optimal manoeuvre strategy, that can be flexible to different mission operation scenarios.
When it comes to the design of a space mission, it is important to define the main operational modes and the transitions among them. For formation flying, one of the main factors that influence the design is the collision-free flight during the overall mission profile. A suitable collision avoidance strategy should be defined among the satellites in the formation \citep{Wermuth2015,koenig2018safe}. For Earth observation missions, the close formation is introduced to achieve the concept of a distributed scientific payload. In the particular case of a potential follow-on to ESA's Soil moisture and Oceans Salinity (SMOS) mission, it would be required for the satellites to fly at a distance in the order of ten meters. Such close separation poses a stringent requirement on inter-satellite collision avoidance for a safe flight. 

The concept of formation flying reconfiguration has been extensively analysed in the literature \citep{tillerson2002co,armellin2004optimal,acikmese2006convex,Morgan2014mpc,koenig2018safe,sarno2020guidance}. 
\cite{acikmese2006convex} proposed a guidance algorithm for formation reconfiguration with heuristic collision avoidance constraints when the satellite distances are in the order of tens of metres. The formulation as a second-order cone algorithm is suitable for onboard implementation. 
\cite{Morgan2014mpc} proposed a decentralised model predictive control algorithm for reconfiguration of swarms of spacecraft in $J_2$ invariant orbits, with a flight distance in the order of hundreds of metres. He proposed a convex approximation of the prohibited zone for a collision-free flight area of the swarms of spacecraft. A similar approach was described by \cite{sarno2020guidance}, who proposed an autonomous low thrust reconfiguration for a distributed system when the relative distances are in the order of hundreds of metres. The path planning was optimised via genetic algorithms and integrated with a convex optimisation law for rapid onboard computation of the control law.
Starting from these literature findings, this work proposes a strategy for reconfiguration that can be applied to close formation flying for remote sensing applications in general, and for Formation Flying L-band Aperture Synthesis (FFLAS) in particular \citep{scala2020formation}. The low-thrust engine technology is considered for the control law implementation, allowing the continuous control of the satellites' state. This is essential for close formation, in the order of few tens of metres, to continuously control the relative position and avoid possible unwanted behaviour, which might lead to a collision. Moreover, a low-thrust technology could provide a good delta-v capacity for a long mission lifetime. For the case of FFLAS, a mission duration of at least 5 years is envisioned. The development of an inter-satellite collision avoidance strategy is presented, to deal with the close proximity of the satellites. The strategy proposed exploits the benefits of the convex optimisation, to find the global optima of the control law, as in \cite{Morgan2014mpc} and \cite{sarno2020guidance}. Additional constraints are introduced in the definition of the problem, to simulate a real mission scenario. A limitation upon the maximum thrust available is introduced to deal with the technological limitation of the onboard engine. 
The algorithm presented is applied to the FFLAS mission concept for remote sensing, which is being developed at Politecnico di Milano in collaboration with Airbus Defence and Space Spain, under a European Space Agency (ESA) mission study \citep{scala2020formation}. The transition among the main mission modes is simulated, to present a delta-v optimal solution for a three-satellite formation flying. In particular, the transition between the nominal Earth observation mode and the Cold Sky Pointing mode is simulated for the calibration of the scientific payload at least once per month. 

The work presented in this paper starts from the literature findings applied to formations baseline in the range of tens to thousands of kilometres. The aim is to extend the approach for fuel-optimal trajectories to close formations with an inter-satellite distance in the order of few tens of meters. To provide an accurate description of the dynamics, the linearised dynamical model is based on ROEs representations, including the effect of the mean Earth oblateness term, $J_2$. 
For a better inclusion of this effect, the relative dynamics is described in the Relative Orbital Elements (ROEs) framework. This representation provides a accurate description of the dynamics, in the order of few centimetres for a time-frame of 10 to 20 orbital periods \citep{gaias2018semi}.
The accuracy of the model with respect to the non-linear description of the dynamics has an important impact on the reliability of the solutions. Being able to provide a trajectory description with a dynamical representation accurate at the centimetres level is of primary importance for more refined analyses, such as the inclusion of the navigation sensors in the state reconstruction.
The paper is organised with an initial overview in Section \ref{sec:Background}, describing the reference systems used in the analysis. Moreover, it presents the relative orbital dynamical model in the ROEs environment, based on a semi-analytical model. The direct and inverse transformation between the relative state in the Hill frame and the ROEs framework is presented, starting from the work by \cite{gaias2020safe}. The remote sensing scenario for Earth observation is presented in Section \ref{sec:ProblemDef}, with the FFLAS mission based on an L-band aperture synthesis interferometer. Then, the optimal formation reconfiguration methodology is described in Section \ref{sec:methodology}. Starting from the classical formulation of the control system, first, a discretisation procedure is implemented, and then, the problem is converted into a convex formulation. The optimal control is described as a fuel-optimal problem, constrained by the maximum thrust level given by the low-thrust engine and by the minimum inter-satellite distance. The convex optimisation problem is solved with the disciplined convex programming, based on the \textit{CVX} software developed by \cite{grant2013cvx}. Finally, Section \ref{sec:study_Case} presents the application of the proposed methodology to the three satellites mission study for remote sensing, FFLAS. The main operational modes are presented and the optimal control law for the transition is proposed for a fast reconfiguration, in less than an orbital period of the reference orbit. Specifically, the transition from the nominal Earth observation mode to both the cold sky pointing and the safe mode is proposed, together with a collision avoidance strategy in case of failure of the main engine of one satellite. 


\section{Reference frames and dynamical model}\label{sec:Background}
This section provides an overview of the reference frames and relative dynamical equations used to develop the manoeuvre strategy. The analysis is carried out in both the relative orbital elements framework and the classical Radial-Transversal-Normal (RTN) frame. The former allows a better representation and inclusion of the main perturbations, such as the Earth oblateness $J_2$, in the relative motion. On the other hand, the RTN frame provides an immediate and straightforward inclusion of the maximum thrust limitation and collision avoidance constraints. 
Section \ref{sub:RefFrame} gives an overview of the absolute orbital frame, the RTN frame, the ROEs framework and the body-fixed frame. The latter allows a representation of the control effort in the body axis, providing information for the design and configuration of the onboard thruster. Section \ref{sub:Relative_motion} presents the State Transition Matrix (STM) in the ROEs framework used to describe the natural free motion of the formation flying. Since the control effort is easily included in the RTN frame, we transformed back the STM into this frame for a better representation of the natural forced motion of the formation.

\subsection{Reference frames} \label{sub:RefFrame}
An accurate definition of the reference systems used to describe the absolute and relative dynamics is of importance to have a clear insight of the analyses performed to define the reconfiguration strategies of the formation flying.
\subsubsection{Inertial absolute orbital frame} \label{sub:OrbitFrame}
The orbital frame used to describe the absolute position and velocity is the Earth Mean Equator at J2000 epoch (EMEJ2000), at midnight. The absolute state vector, including position and velocity, can be defined in the EMEJ2000 for both the reference orbit and the satellites orbit. In this paper, the reference orbit is identified with subscript $c$ in the state vector:
\begin{equation}
\mathbf{x}_c = \left\{ \begin{array}{c}
\mathbf{r}_c\\ 
\mathbf{v}_c
\end{array} \right\}
\end{equation}
with $\mathbf{r}_c$ and $\mathbf{v}_c$ being the position and velocity vectors of the reference orbit respectively. Considering a formation flying of $N$ satellites, the absolute state of a generic $j$ spacecraft, with $j=1,...,N$, is defined as:
\begin{equation}
\mathbf{x}_j = \left\{ \begin{array}{c}
\mathbf{r}_j\\ 
\mathbf{v}_j
\end{array} \right\}
\end{equation}
with $\mathbf{r}_j$ and $\mathbf{v}_j$ being the position and velocity vectors of the $j$-th orbit respectively. From the absolute state vector $\mathbf{x}$, it is possible to represent the Keplerian orbital elements of the reference orbit and the formation flying satellites. In this paper, we consider the mean Keplerian elements, instead of the classical osculating ones. 
We define the mean Keplerian elements of the reference orbit as $\alpha_c = \{a_c,\,e_c,\,i_c,\,\Omega_c,\,\omega_c,\,M_c\}$, with the semi-major axis, the eccentricity, the inclination,  the right ascension of the ascending node, the argument of perigee, and the mean anomaly, respectively. Similarly for a generic satellite $j$ in the formation, we define the mean Keplerian elements as $\alpha_j = \{a,\,e,\,i,\,\Omega,\, \omega,\,M \}_j$ for $j=1:N$.

\subsubsection{Radial-transversal-normal frame}
\begin{figure}[htb]
	\centering
	\includegraphics[width=0.30\textwidth]{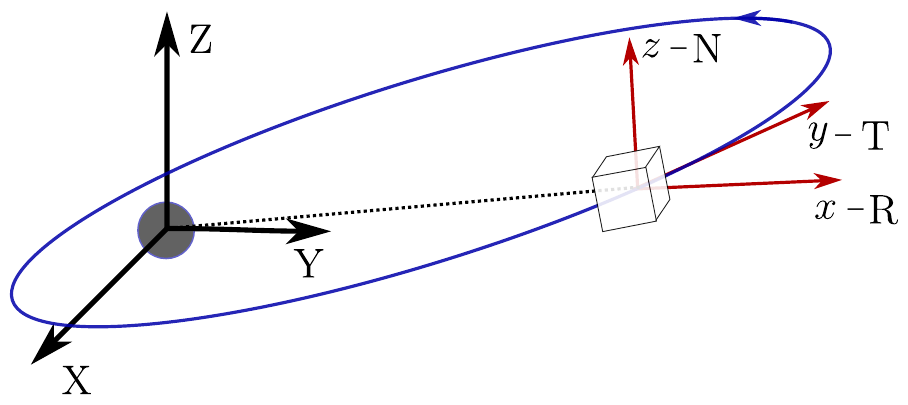}
	\caption{RTN frame centred in the formation flying reference orbit.}
	\label{fig:RTN_frame}
\end{figure}

The relative motion is typically described in the Radial-Transversal-Normal orbital frame, also defined as the Hill Orbital Frame, introduced by \cite{hill1878researches}. The RTN frame is commonly used in the representation of the formation flying relative motion, due to the insight representation of the time evolution of the relative satellites’ position, \citep{d2005relative}. The RTN unit vector triad is defined as $e_R$, $e_T$ and $e_N$, where: $e_R$ is aligned with the radial direction, pointing outward, $e_N$ is aligned with the orbit momentum vector, $e_T$ to complete the right-hand coordinates system.
Their mathematical expressions are reported in Eq.~(\ref{eq:RTN_frame}), with $r_c$ and $v_c$ the inertial position and velocity of the reference orbit.
\begin{equation}
\label{eq:RTN_frame}
\begin{array}{l}
\mathbf{e_R} = \mathbf{r}_c/r_c\\[0.3cm] 
\mathbf{e_N} = (\mathbf{r}_c \times \mathbf{v}_c)/\|\mathbf{r}_c\times \mathbf{v}_c\|\\[0.3cm] 
\mathbf{e_T} = \mathbf{e_N}\times\mathbf{e_R}
\end{array} 
\end{equation}
This frame is described by a rotation of the synodic frame given by the mean motion of the reference orbit $\mathbf{\omega}=n\, \mathbf{e_N}$, where $n$ is the mean motion. The RTN frame is represented in Fig.~\ref{fig:RTN_frame}, where the case of a quasi-circular orbit is represented, with the transversal axis in the direction of the orbital velocity.

\subsubsection{Relative orbital elements frame}
The relative orbital elements framework was described by \cite{d2005relative} for easier inclusion of the orbital perturbation in the modelling of the relative motion. Differently from the classical RTN representation, the ROEs allows a semi-analytical representation of the dynamical model, with a better insight into the relative motion. Moreover, the resulting control model accuracy is higher than the classical RTN approach \citep{gaias2020analytical}.
The ROEs describe the orbital elements of each satellite in the formation with respect to their reference orbit. Starting from the mean Keplerian elements of the reference orbit and a generic satellite $j$ of the formation as $\{a_c,\,e_c,\,i_c,\,\Omega_c,\,\omega_c,\,M_c\}$ and $\{a,\,e,\,i,\,\Omega,\,\omega,\,M\}_j$, respectively, as defined in Section \ref{sub:OrbitFrame}, the ROEs can be computed.  The relative orbital elements of the $j$-th satellite are defined in Eq.~(\ref{eq:ROEs_def}) by $\delta \bm{\alpha}_j$, as a set of dimensionless variables from the formulation in \cite{d2005relative}.

\begin{equation}
\label{eq:ROEs_def}
\delta\bm{\alpha}_j = \left\{ \begin{array}{c}
\delta a \\
\delta \lambda \\ 
\delta e_x \\ 
\delta e_y \\ 
\delta i_x\\ 
\delta i_y
\end{array}\right\}_j  = \left\{\begin{array}{c}
(a - a_c)/a_c \\
u - u_c + (\Omega - \Omega_c)\cos(i_c) \\
e \cos(\omega) - e_c \cos(\omega_c)\\
e \sin(\omega) - e_c \sin(\omega_c) \\
i - i_c \\
(\Omega - \Omega_c) \sin(i_c)
\end{array}\right\}_j,
\end{equation}

\begin{figure}[htb]
	\centering
	\includegraphics[width=0.45\textwidth]{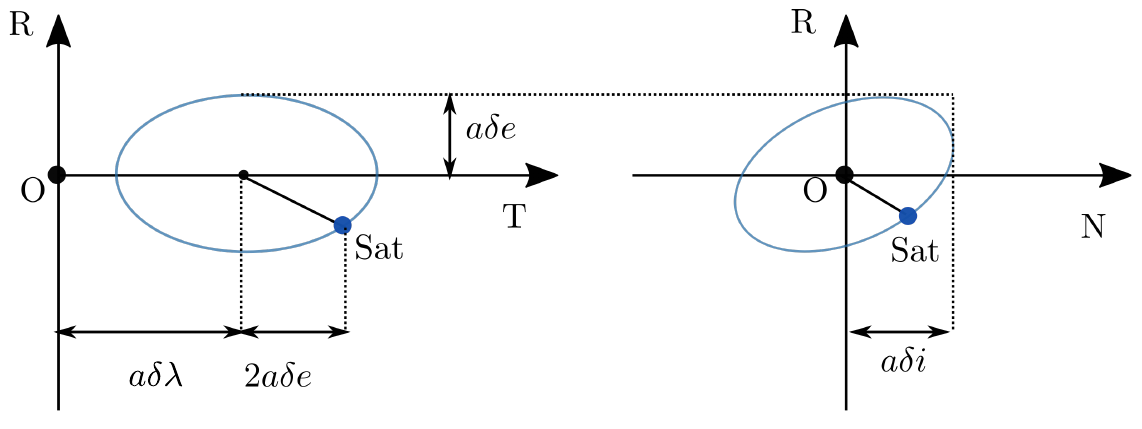}
	\caption{Interpretation of relative orbital elements in the RTN frame.}
	\label{fig:ROEs_interpretation}
\end{figure}

Where $u = M + \omega$ is the mean argument of latitude, depending on the mean anomaly $M$ and on the argument of perigee $\omega$. The ROEs are composed of the relative semi-major axis $\delta a$, the relative mean longitude $\delta \lambda$ and the relative eccentricity $\delta e$ and inclination  $\delta i$ vectors respectively. The interpretation of the ROEs in the RTN frame is shown in Fig.~\ref{fig:ROEs_interpretation}, where the ROEs are shown in terms of cross-track and along-track displacement. 

\subsubsection{Body-fixed frame}\label{sec:body_frame}
\begin{figure}[htb]
	\centering
	\includegraphics[width=0.30\textwidth]{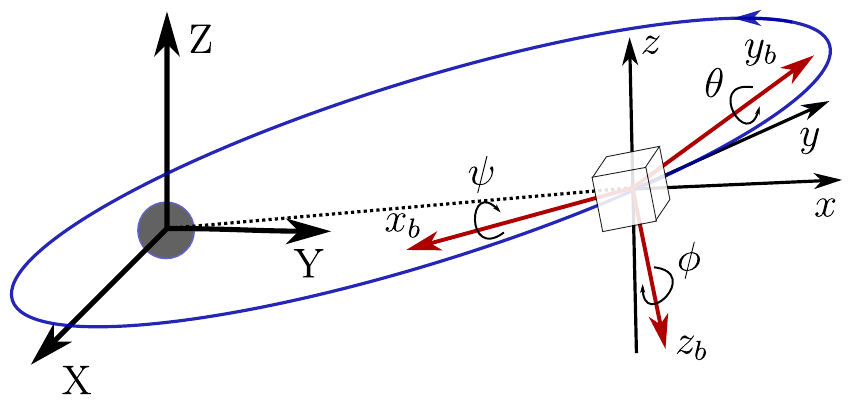}
	\caption{Satellite body-fixed reference frame.}
	\label{fig:body_fixed_frame}
\end{figure}

A third reference frame that is considered in the analysis is the body-fixed frame. It allows the representation of the $j$-th spacecraft attitude profile in a frame that is tied to the satellites' system. This reference frame is centred in the satellite body (centre of mass) and follows the attitude evolution in time. 
There is no standard convention for defining this frame, but it is typically selected along the symmetry axis and/or the main axes of inertia \citep{montenbruck2015gnss}. In this paper, the body-fixed frame $\{x_b,\,y_b,\,z_b\}$ is defined following the principal axis of the satellites, and we consider the three-axis of the on-board engine aligned with this frame, to enable a consistent description of the control effort of the $j$-th satellite. 
It is important to describe the transformation among the body-fixed frame and the RTN frame. This is typically described by three subsequent rotation of the yaw, roll and pitch angle ($\psi$, $\theta$, and $\phi$ respectively), as shown in Fig.~\ref{fig:body_fixed_frame}. The rotation matrix that relates the RTN and the body fixed frame is $ \mathbf{R}_{b}^{RTN} = \mathbf{R}_{z_b}(\phi)\cdot \mathbf{R}_{y_b}(\theta) \cdot \mathbf{R}_{x_b}(\psi)$, where $\mathbf{R}$ is the rotation matrix around the $x_b,\,y_b$ and $z_b$ axis of an angle $\phi$, $\theta$ or $\psi$ respectively. The matrix $ \mathbf{R}_{b}^{RTN}$ is used for the conversion between the state vector in the RTN frame ($\mathbf{x}_{RTN}$) and the in that body axis ($\mathbf{x}_{b}$):
\begin{equation}
\label{eq:RTN_to_body}
\left\{\begin{array}{c}
\mathbf{x}_b = \mathbf{R}_{b}^{RTN} \cdot \mathbf{x}_{RTN}\\ 
\mathbf{x}_{RTN} = (\mathbf{R}_{b}^{RTN})^T \cdot \mathbf{x}_b
\end{array} 
\right.
\end{equation}

For a remote sensing formation flying mission, such as FFLAS, the nominal attitude of the satellites is Earth pointing for observation purposes. This condition requires the $x_b$ axis aligned with the radial direction, but with the opposite sign (pointing to the centre of the Earth). In this work we assume that the attitude of the satellites during the manoeuvres can ensure the alignment of the $y_b$ and $z_b$ axes with the transversal and normal directions respectively, maintaining the thrusters in the correct direction.

\subsection{Relative dynamical model} \label{sub:Relative_motion}

This work focuses on the dynamical representation of relative motion in the LEO environment, at an altitude higher than 500 km (see Section \ref{sec:ProblemDef}).  The most relevant perturbing effect for such a condition is given by the Earth's oblateness, while the other sources are considered negligible in the preliminary analysis presented in this paper. The solar radiation pressure and the third body perturbation are at least an order of magnitude lower than the primary $J_2$ effect. 
In literature, the use of the Hill-Clohessy-Wiltshire (HCW) equations has been widely discussed. The classical HCW equations are typically used to describe the natural relative motion, without the inclusion of external orbital perturbations \citep{clohessy1960terminal}. The orbital perturbations, like the Earth's oblateness or the atmospheric drag, can highly influence the natural motion in LEO orbits. Thus, the HCW equations can introduce significant errors in the long-term prediction of the relative motion of satellites. To account for these limitations, the main perturbation effects have been included in several models in the RTN reference frame for a quasi-circular reference orbit. \cite{vadali2000hill}, \cite{schweighart2002high} and \cite{izzo2003new} proposed a natural relative motion model, including the Earth's oblateness' effect. A similar model was developed by \cite{Sabatini2008}, including also the drag disturbance. In their work, they demonstrated how the proposed model provided a dynamical behaviour closer to the non-linear propagator, with respect to the previous ones. Thus, the thrust free relative motion for the $j$-th satellite in the formation could be expressed in the RTN frame by the following relation:
\begin{equation}
\label{eq:rtn_dyn_sys}
\dot{\mathbf{x}}_j(t) = \mathbf{A}_{rtn}(t) \,\mathbf{x}_j(t)
\end{equation}
where, $\mathbf{A}_{rtn}(t)$ is the state matrix including the Earth's oblateness effect as described in detailed in \cite{Sabatini2008}, and $\mathbf{x}_j(t) = \left\{ x,\,y,\,z,\,\dot{x},\,\dot{y},\,\dot{z}\right\}'$ is the state vector of the $j$-th satellite. For the sake of this work, the drag effect is not considered, since the satellites in the formation are close to each other, with an inter-satellite distance in the order of few tens of meters, and the effect of the differential effect of the drag could be neglect in this preliminary analysis. The model proposed by \cite{Sabatini2008} provides an error of the relative position of about 5\% to 10\% of the relative distance, considering a relative orbit of a few hundred to a few a thousand metres. 

Another approach exploiting the ROEs framework was presented in a previous work by \cite{gaias2018semi}. Only the terms related to the second order zonal contribution $J_2$ are considered in this work, and the first order relative mean motion in ROEs is expressed as:
\begin{equation}
\label{eq:roes_dyn_sys}
\delta\dot{\bm{\alpha}}_j (t) = \mathbf{A}_\alpha (t) \, \delta \bm{\alpha}_j (t)
\end{equation}
where, $\mathbf{A}_\alpha (t)$ is the first-order state matrix, whose terms are reported in the \ref{appendixA} for sake of clarity, and $\delta \bm{\alpha}_j (t)$ are the relative orbital elements of the $j$-th satellite. 
This first-order approach in the $\delta\bm{\alpha}$ variables requires
only small $\delta a$, $\delta e_x$, $\delta e_y$, and $\delta i_x$. The formation flying geometry considered in this work consists of satellites flying on similar quasi-circular orbits, with a small difference in the relative orbital elements. In \cite{gaias2018semi} it is described how this approach provides a small error for the propagation of the ROEs than the classical integration procedure (non-linear approach). For a relative orbit of a few hundreds of meters, the errors remain in the order of a few centimetres to one metre for the first 20 orbital revolutions. To gain such accuracy, it is important to exploit an osculating-to-mean transformation in the same order of the reference dynamics. In particular, since we considered a first-order relation in Eq.~\ref{eq:roes_dyn_sys}, a first-order correction based on both canonical and Lie transformation was used, as described in \cite{gaias2020analytical}.
For a close-range formation – when the inter-satellite separation is less than 1 kilometre – a linear mapping from ROEs to relative states in RTN frame is considered with the procedure presented in \cite{gaias2020safe}. The change of variables is obtained by introducing the Lyapunov transformation \citep{gaias2020safe}.  It requires the knowledge of the mean argument of latitude $u_c$ and of the mean motion $n_c$ of the reference orbit. The matrix used for the mapping from ROEs to relative states in RTN is reported in Eq.~(\ref{eq:roes_to_rtn}). This formulation is non-singular for all time $t$ and it can be inverted with the inverse transformation from the relative state in RTN to ROEs. 
\begin{equation}
\label{eq:roes_to_rtn}
\mathbf{T}(t) = \left[
\begin{array}{cccccc}
1	& 0 & -\cos nt & - \sin nt & 0 & 0 \\
0	& 1 & 2\sin nt & -2\cos nt & 0 & 0 \\
0	& 0 & 0 & 0 & \sin nt & -\cos nt \\
0	& 0 & n\sin nt & -n \cos nt & 0 & 0 \\
-\frac{3\,n}{2}	& 0 & 2n \cos nt & -2\sin nt & 0 & 0 \\
0	& 0 & 0 & 0 & n \cos nt & n \sin nt
\end{array}
\right]
\end{equation}
The direct and the inverse transformations are the following:
\begin{equation}
\label{eq:direct_invers_T_transf}
\left\{\begin{array}{l}
\mathbf{x} = \mathbf{T}(t) \cdot a\delta\bm{\alpha}	\\
a\delta\bm{\alpha} = \mathbf{T}^{-1}(t)\,\mathbf{x}	
\end{array}
\right.
\end{equation}
where $\delta \alpha$ represents the relative orbital elements, and $\mathbf{x}$ the relative states in RTN. These equations can be used to pass from the representation in ROEs, as in Eq.~(\ref{eq:roes_dyn_sys}), to the relative state in RTN representation, and vice versa. The direct transformation from  Eq.~(\ref{eq:roes_dyn_sys}), exploits the change of variables: 
\begin{equation}
\label{eq:rtn_sys_dyn_from_roe}
\dot{\mathbf{x}}_j(t) = \left(\mathbf{T}\mathbf{A}_\alpha(t) \mathbf{T}^{-1} - \mathbf{T}\mathbf{T}^{-1} \right)\mathbf{x}_j = \hat{\mathbf{A}} \mathbf{x}_j(t)
\end{equation}
where $\hat{\mathbf{A}}$ is the new state matrix in RTN for the dynamic motion description. A similar procedure can be done for the inverse transformation starting from Eq.~(\ref{eq:rtn_dyn_sys}).
The representation in Eq.~(\ref{eq:rtn_sys_dyn_from_roe}) maintain the accuracy level of the ROE-based model: the error with respect to the non-linear representation is less than one metre. Its transformation in the RTN frame could still provide such accuracy when the inter-satellite distance is less than one kilometre, thus can be applied for proximity manoeuvre design.

\section{Problem definition}
\label{sec:ProblemDef}
The analyses presented in this work are developed for the three-satellite FFLAS mission, but are valid for multiple formation satellites for remote sensing purposes, in LEO, at a mean orbital altitude of 770 km. The beneficial effect of exploiting formation flying to increase the virtual aperture of the instruments was widely studied \citep{krieger2007tandem,moreira2015tandem}. A close formation, with an inter-satellite distance in the order of tens of meters, can significantly improve the resolution of a Synthetic Aperture Radiometer instrument: the equivalent payload aperture size increases with the dimensions of the formation. Similarly, this work considers the possibility of performing L-band aperture synthesis radiometry using three spacecraft (FFLAS mission concept). Each individual hexagonal array interferometer, as in \citep{kerrLband,Neira2020}, could provide up to 25 km spatial resolution for a payload aperture size of about 15 m diameters. Introducing the formation flight concept, a triangular formation of three arrays with dimension side in the order of 10 m to 15 m, could provide an equivalent aperture of 21 m diameter and up to 10 km resolution \citep{Zurita2013,Neira2020}. This improvement of achieving a larger aperture with spacecraft in formation flight is in the direction of Earth's observations scientific needs, where most of the hydrological processes occur in the 1-10 km scale. Future mission studies for Earth observation could take advantage of the formation flying to improve the spatial resolution and as a consequence the scientific performances capabilities, such as the mission study FFLAS that we are considering in this work as the study case. 
These premises serve as a starting point for the reconfiguration manoeuvres design. As described in Section \ref{sec:Introduction}, some research has already been proposed to study optimal manoeuvre for formation flying \citep{sabatini2009collective,Morgan2014mpc,sarno2020guidance}. Starting from the work done in literature, we propose and apply an optimal reconfiguration method to close formation flying, with the focus on the most relevant manoeuvre required during mission operations. 
To ensure the performance increase exploiting distributed aperture synthesis payloads, it is required that the satellites fly in close formation with a fixed relative position. Specifically, this translates into the need for a forced relative motion during the nominal phase of the mission. In this way, the aperture synthesis can be realised as a single augmented instrument. The low thrust technology is the most suitable to continuously control the relative position among the satellites. Moreover, the instruments should be at the same orbit altitude to ensure correct interferometric processing. For this reason, a formation flight lying on the transversal-normal plane (Fig.~\ref{fig:RTN_frame})  enables such performances. Another important consideration concerns the relative inter-satellite distance. Depending on the array geometry of the payload, it can be shown that a proper selection of the relative distance is required to perform interferometry \citep{Zurita2013}. Considering an L-band array with a hexagonal geometry, the best formation for interferometry is obtained by rigidly flying three hexagons facing arms at the vertices of an equilateral triangle of side equal to twice the inner diameter of the hexagon. These considerations result in a close - tight - formation, where the small inter-satellite distance is a critical aspect and requires a robust collision avoidance strategy.
A final aspect that should be considered for the FFLAS mission is the need for periodic calibration of the synthetic aperture payload. The instrument should be typically calibrated once a month, pointing the instrument in the cold sky direction. This translates into the need of designing a fuel optimal manoeuvre for the transition between the nominal Earth Pointing Mode (EPM) and the Cold Sky Pointing Mode (CSPM), for calibration purposes. 
The previous considerations introduce some requirements for the satellite formation establishment of FFLAS.  The close - tight - formation requires a forced relative motion to control the relative position, where the low-thrust technology is the most suitable to provide continuous control, even with a limitation in the maximum thrust level. Moreover, for observation purposes, the aperture plane of the formation will lie in the transversal-normal plane, and then the satellites will perform a calibration manoeuvre at least once per month, to switch from EPM to CSPM configuration. Finally, an inter-satellite collision avoidance strategy should be implemented in the manoeuvre, to tackle the close relative position among the satellites.


\section{Orbital control methodology} \label{sec:methodology}
This section presents the model and the methodology followed to design optimal low-thrust manoeuvres for FFLAS, or in general, remote sensing close formation flight.  
For this work, the initial and final state vectors of the satellites depend on the mission operations profile.  
Thus, the optimal formation reconfiguration strategy relies on the knowledge of initial and final condition a priori. The degree of freedom in this sense is the time of the manoeuvre, which can be set as a fraction/multiple of the orbital period to compute the manoeuvre trajectory and the control terms. Both the constraints on the maximum available thrust and the collision avoidance are considered in the implementation of the optimal control problem.

The analysis presented in this section is based on the transformation of the classical optimal control problem into a convex formulation. The convexification of the problem grants the existence of a unique solution of the problem, and it does not require several iterations to converge to the optimal solution \citep{Morgan2014mpc}. An approach, relying on convex formulation, has already been discussed in previous studies for relative motion manoeuvre  \citep{mitani2013continuous,Morgan2014mpc,koenig2020fast,sarno2020guidance}. Moreover, \cite{scala2020three} applies the convex approach for formation reconfiguration after the satellite injection by the launcher. The convex representation allows a simple immediate discretisation of the system dynamics, including the effect of the control action. The reduced computational effort required by this approach \citep{Morgan2014mpc}, is suitable for an implementation on-board the satellites, which have reduced computational capability. The solution of the convex optimisation problem provides the thrust level for the guidance of the optimal reconfiguration manoeuvre of the spacecraft. 

The reconfiguration manoeuvre can be described as a global time-optimal or a fuel-optimal problem. In this case, the approach was to fix the time for the reconfiguration and optimise propellant consumption. Now, the optimal problem is manipulated into a convex form and this requires the discretisation of the entire control problem, for both constraints and cost function.

\subsection{Optimal control system}
\label{sec:OCP}
The fuel-optimal reconfiguration manoeuvre can be defined under the classical Optimal Control Problem (OCP) formalism. 
The control system is described by an ordinary differential equation for the $j$-th satellite in the formation.
\begin{equation}
\label{eq:OCS_dyn}
\dot{\mathbf{x}}_j(t) = f\left(t,\mathbf{x}_j,\mathbf{u}\right) = \mathcal{A}_j \mathbf{x}_j(t) + \mathcal{B}_j \mathbf{u}_j(t) 
\end{equation}
where $\mathbf{x}_j = \left\{x_j, \, y_j,\, z_j,\, \dot{x}_j,\,\dot{y}_j,\, \dot{z}_j \right\}^T$ is the state corresponding to the $j$-th satellites state vector, $\mathbf{u}_j = \left\{u_{x_j},\, u_{y_j}, \, u_{z_j} \right\}^T$ is the control input vector , and $t$ is the time. The matrix $\mathcal{A}$ is the matrix representing the natural relative motion under the influence of external perturbations, and $\mathcal{B}$ is the control matrix. The low thrust control is introduced in the dynamical system as a continuous effect on the natural dynamics. This paper describes the control in the RTN relative frame as a control input $\mathbf{u}(t)$, included in the dynamical system through the control input matrix $\mathcal{B}$:
\begin{equation}
\mathcal{B} = \left\{\begin{array}{ccc}
0 & 0& 0 \\
0& 0 & 0 \\
0 & 0& 0 \\
1& 0& 0 \\
0& 1& 0 \\
0& 0& 1
\end{array} 
\right\}
\end{equation}
The matrix $\mathcal{B}$ relates the control term to the velocity components in the system dynamics. The objective of the analysis is to find the optimal control input $\mathbf{u}(t)$, such that the performance index, or cost function, is minimised. The performance index for this fuel-optimal control problem is defined for each $j$-th satellite as following:
\begin{equation}
\label{eq:performance_index}
J = \int_{t} \left\|\mathbf{u}_j(t) \right\|_{1} dt
\end{equation}
The 1-norm is used in the cost function to minimise the sum of the magnitude of the control components in the RTN directions. This corresponds to the minimisation of the propellant mass for the manoeuvre; the control effort is related to the propellant mass via the spacecraft wet mass and the engine thrust level: $\mathbf{T}_j = m_{s/c_j} \cdot \mathbf{u}_j $. 
The initial and final conditions of the dynamic system in Eq.~(\ref{eq:OCS_dyn})  influence the dynamics with the following relations:
\begin{equation}
\label{eq:IC_FC}
\left\{\begin{array}{l}
\mathbf{x}_j(t_0)=\mathbf{x}_{0,j}\\ 
\\
\mathbf{x}_j(t_f)=\mathbf{x}_{f,j}
\end{array} \right.
\end{equation}
where $\mathbf{x}_{0,j}$ is the initial state of the $j-th$ satellite and $\mathbf{x}_{f,j}$ is the final (or boundary) condition. Finally, as discussed in Section \ref{sec:ProblemDef}, the problem is subject to some constraints on the collision avoidance and the maximum available thrust. Thus, at any time instant, the maximum thrust limitation translates into a limitation in the maximum acceleration possible $T_{max} = m_{s/c_j} \cdot \mathbf{a}_{max_j} $. The collision avoidance constraint is expressed in terms of minimum allowable distance between the $j$-th and $i$-th satellites. 
\begin{equation}
\label{eq:Constraints}
\left\{\begin{array}{l}
\left\|\mathbf{u}_j(t)\right\|\leq\mathbf{a_{max}}_j\\ 
\\
\left\| \mathbf{C}\left(\mathbf{x}_j(t) - \mathbf{x}_i(t)\right) \right\|_2 \geq d_{thr}
\end{array} \right.
\end{equation}
where $\mathbf{a_{max}}_j$ is the maximum acceleration that the thruster can provide, $\mathbf{C} = \left[\mathbf{I}_{3\times 3}\quad \mathbf{0}_{3\times 3}\right]$ is the matrix to retain only the position components in the state vectors $\mathbf{x}_j(t)$ and $\mathbf{x}_i(t)$, and $d_{thr}$ is the minimum safe distance to avoid inter-satellite collision. 

\subsection{Convex Optimal Control System}
In this section, the discretisation procedure for the OCP is described. The discretisation procedure is the first step to write an optimal control problem in a convex form. 
This is conceptually similar to the least-square and the linear programming. A Convex Optimal Control Problem (COCP) can be efficiently solved with different methods, such as the interior-point methods \citep{boyd2004convex}. Thanks to the sparsity of the matrices involved, it is typically very time efficient and can handle large problems up to thousands of variables and constraints. The most challenging part to deal with the convex formulation is the transformation of the OCP into a convex form. Three main requirements define a convex problem: both the objective and the inequality constraint functions must be convex, and the equality constraints must be affine \citep{boyd2004convex}. The main advantage of a COCP is the equivalence among local and global optimal points.
The first step to convert an OCP in classical form into a convex formulation is the discretisation of both the objective and the constraint functions, in Eqs. (\ref{eq:OCS_dyn}) to (\ref{eq:Constraints}).

\subsubsection{Discretisation procedure}
\label{sec:discretisation}
The approach followed for discretising the system is based on the Laplace transformation of the state equations \citep{rowell2002state}. The time is divided into $K$ finite time instants, each representing the sample interval for the state $\mathbf{x}$ and the update interval for the control term $\mathbf{u}$. Moreover, the zero-order-hold approach is considered, with the control term piecewise constant in each time instant $k$ \citep{Morgan2014mpc}. For the procedure, we considered:
\begin{itemize}
	\item Time discretisation: $k = 1,...,K$ ( where $t_{k=1}=t_0$ and $t_{k=K}=t_f$)
	\item Time interval: $\Delta t = t_{k+1} - t_{k}$
	\item Total time: $T = K-1 \Delta t$
	\item Number of satellites in the formation: $j = 1,..., N$
\end{itemize}
\paragraph{System dynamics}
The discretisation of Eq.~(\ref{eq:OCS_dyn}) requires the solution of the non-homogeneous system, via the Laplace transformation. This procedure leads to the following expression, which includes the convolution integral for the control effort \citep{acikmese2006convex}. 
\begin{equation}
\mathbf{x}_j[k+1] = e^{\mathcal{A}\Delta t} \, \mathbf{x}_j[k] + \int_{0}^{\Delta t} e^{\mathcal{A}\tau}d\tau \, \mathcal{B}\, \mathbf{u}_j[k]
\end{equation}
Now, considering that the state matrix $\mathcal{A}$ is invertible, the integral term is expressed as:
\begin{equation}
\int_{0}^{\Delta t} e^{\mathcal{A}\tau}d\tau = \mathcal{A}^{-1}\left(e^{\mathcal{A}\Delta t} - \mathbf{I}\right) 
\end{equation}
where $\mathbf{I}$ is the identity matrix. Finally, recalling that the expression $e^{\mathcal{A}\Delta t}$ is equivalent to the matrix formulation $\mathbf{I} + \mathcal{A}\Delta t$, we obtain the final expression of the system dynamics:
\begin{equation}
\label{eq:discrete_sys_dyn}
\mathbf{x}_j[k+1] = \left(\mathbf{I} + \mathcal{A}\Delta t\right) \, \mathbf{x}_j[k] + \mathcal{B}\, \Delta t\, \mathbf{u}_j[k] 
\end{equation}
where $j=1,...,N$. The matrix $\mathcal{A}$ represents the system dynamics of the formation flying.  Depending on the nominal value of the inter-satellite distance, different considerations should be done.
		First, for an inter-satellite distance $\leq 1\, km$, the linear mapping between ROEs and RTN frame reported in Eq.~(\ref{eq:roes_to_rtn}) is valid and the ROE-based state matrix of Eq.~(\ref{eq:rtn_sys_dyn_from_roe}) can be used. For this case of proximity operations, when the matrix $\mathcal{A}$ correspond to the state matrix of Eq.~(\ref{eq:rtn_sys_dyn_from_roe}), the description in relative orbital elements is considered to maintain a better accuracy level in the optimal manoeuvre design. This case could be applied to the nominal operations required by a remote sensing mission, like FFLAS, such as the payload calibration manoeuvre or the safe mode transition.
		On the other hand, for different mission operational scenarios, the inter-satellite distance could be larger than 1 km, such as after the orbital injection by the launcher and the subsequent initial reconfiguration manoeuvre to set the nominal formation geometry. For this case, the linear mapping in Eq.~(\ref{eq:roes_to_rtn}) is not accurate and the formation dynamics could not benefit from the relation in Eq.~(\ref{eq:rtn_sys_dyn_from_roe}). In this case, the dynamic is described by the RTN-based state matrix, as in the formulation of Eq.~(\ref{eq:rtn_dyn_sys}) \citep{Sabatini2008}.

\paragraph{Objective function} The objective function (or cost function) described in Eq.~(\ref{eq:performance_index}) is dicretised thanks to the piecewise constant control property in each time interval:
\begin{equation}
J = \sum_{1}^{K} \left\| \mathbf{u}_j [k]\right\|_1 
\end{equation}
where $j=1,...,N$ and the 1-norm is used again for an optimal solution with minimum fuel consumption. 

\paragraph{Initial and final conditions} The initial and final conditions of the system dynamics can be described as:
\begin{equation}
\label{eq:IC_FC_discrete}
\left\{\begin{array}{l}
\mathbf{x}_j[k=1]=\mathbf{x}_{0,j}\\ 
\\
\mathbf{x}_j[k=K]=\mathbf{x}_{f,j}
\end{array} \right.
\end{equation}
where the relation is valid for each satellite in the formation with $j = 1,...,N$.

\paragraph{Thrust constraints} The maximum thrust limitation constraint in Eq.~(\ref{eq:Constraints}) can be discretised as following:
\begin{equation}
\label{eq:thrust_lim_discrete}
\left\|\mathbf{u}_j[k]\right\|\leq a_{max_j} 
\end{equation}
where $j=1,...,N$ and the maximum bound is imposed so that at each time interval, the acceleration provided by the thrusters is bounded by the engine technological limit. 

\paragraph{Inter-satellite collision avoidance constraint}
The minimum allowable distance between the $j$-th an $i$-th satellites (with $j\neq i$) requires a mode detailed discussion. First, the expression in Eq.~(\ref{eq:Constraints}) is discretised as follows:
\begin{equation}
\label{eq:CA_discrete}
\left\| \mathbf{C}\left(\mathbf{x}_j[k] - \mathbf{x}_i[k]\right) \right\|_2 \geq d_{thr}
\end{equation}
where $j=1,...,N-1$ and $i>j$. To guarantee that the collision avoidance constraint is satisfied on $(t_k,\,t_{k+1})\, \forall k$, and to prepare the relation for a correct convexification, the expression in Eq.~(\ref{eq:CA_discrete}) is transformed in the following relation \citep{Morgan2014mpc}:
\begin{equation}
\label{eq:coll_avoidance_convex_final}
\left(\bar{\mathbf{x}}_j[k] - \bar{\mathbf{x}}_i[k]\right)^T\mathbf{C}^T\mathbf{C} \left(\mathbf{x}_j[k] - \mathbf{x}_i[k]\right) \geq d_{thr} \left\|\mathbf{C}\left(\bar{\mathbf{x}}_j[k] - \bar{\mathbf{x}}_i[k]\right)\right\|_2
\end{equation}
for $i>j$ and $j=1,...N-1$, where the $\bar{\mathbf{x}}_j[k]$ represents an initial guess of the optimal trajectory followed by the spacecraft. The closer the initial guess is to the actual trajectory, the more accurate the convex program solution will be. In this work, the initial guess is selected as the result obtained from a first running of the convex problem, without collision avoidance constraints. Then, the collision avoidance is added to the problem formulation and the initial guess is refined with the resulting trajectory and control from the first running of the convex problem. Finally, this refined initial condition is used to obtain a refined solution.
The expression in Eq.~(\ref{eq:coll_avoidance_convex_final}) generates separating planes among the satellites, transforming the circular prohibited zone of Eq.~(\ref{eq:CA_discrete}) into a suitable convex formulation. This formulation defines a collision-free zone with separating planes, ensuring a sufficient condition for the collision avoidance of the $j$-th and $i$-th satellite couple \citep{Morgan2014mpc}.

\subsubsection{Convex optimal control system}
The classical formulation of the control system was discretised in Section \ref{sec:discretisation}, and it was prepared for the convex formulation. Now, using the approach described in \cite{boyd2004convex}, we want to express the control system in a convex formulation, with equality and inequality constraints:
\begin{equation}\label{eq:convex_template}
\begin{array}{lll}
\mathrm{minimise:}& f_0(x) &  \\
\mathrm{subject\, to:}& f_i(x)\leq 0 & i=1,...,m  \\
& h_i(x) = 0 & i=1,...,p
\end{array}
\end{equation} 
where $x\in \mathbf{R}^n$ is the optimisation variable, including the state vectors of all the satellites in the formation. The aim is to solve the optimisation problem for the overall formation to properly minimise propellant while dealing with the inter-satellite collision avoidance constraint.

For each satellite $j$, with $j=1,...,N$, we can define a column vector $\hat{\mathbf{x}}_j$, which includes the state vector and the control term at each time instant $k$:
\begin{equation}\label{eq:jth_state_controlo_vector}
\hat{\mathbf{x}}_j = \left\{x^1_j \quad \cdots \quad x^k_j \quad \cdots \quad x^K_j \quad u^1_j \quad \cdots \quad u^k_j \quad \cdots \quad u_j^{K-1}\right\}^T
\end{equation}
where $\hat{\mathbf{x}}_j$ is a $(6K + 3 (K-1))$ size vector. For conciseness, we define $M = (6K + 3 (K-1))$ the length of the decisional vector of each $j$-th satellite. For the whole formation, we define the full state column vector $\hat{\mathbf{\mathcal{X}}}$ as the decisional vector, with size $(N\cdot M)$:
\begin{equation}\label{eq:full_state_controlo_vector}
\hat{\mathbf{\mathcal{X}}} = \left\{\begin{array}{c}
\hat{\mathbf{x}}_1	\\
\vdots	\\
\hat{\mathbf{x}}_j	\\
\vdots	\\
\hat{\mathbf{x}}_N	
\end{array}
\right\}
\end{equation}
\paragraph{System dynamics} The relation in Eq.~(\ref{eq:discrete_sys_dyn}) is now expressed in terms of the full state column vector $\hat{\mathbf{\mathcal{X}}}$. For each $j$-th satellite, the discrete system dynamic at instant $k$ is:
\begin{equation}\label{eq:discr_sys_dyn_1}
\mathbf{x}_j^{k+1} - \left(\mathbf{I} + \mathcal{A}\Delta t\right) \, \mathbf{x}_j^k - \mathcal{B}\, \Delta t\, \mathbf{u}_j^k = 0
\end{equation}
where $j=1,...,N$. Considering the $k$-th instant, in matrix form the system dynamics for satellite $j$ are the following.
\begin{equation}
\label{eq:convex_sys_dyn}
[\mathbf{0}_{6\times6(k-1)}, \quad -(\mathbf{I}_{6} + \mathcal{A}\Delta t), \quad \mathbf{I}_6, \quad \mathbf{0}_{6\times 3(2K - k -3)}, \quad - \mathcal{B}\Delta t, \quad \mathbf{0}_{6\times 3(K-k-1)} 
] \cdot \hat{\mathbf{x}}_j = 0
\end{equation}
where the matrix is called $\mathbf{\mathcal{A}}_{sd}$. Thus, for the overall formation, the system dynamics can be expressed as: 
\begin{equation}\label{eq:FF_convex_sys_dyn}
\left[\begin{array}{ccc}
\cdots	& \cdots & \cdots  \\
\mathbf{0}_{6(K-1)\times M(j-1)} & \mathcal{A}_{sd} & \mathbf{0}_{6(K-1)\times M(N-j)} \\
\cdots	& \cdots & \cdots 
\end{array}
\right] \cdot \hat{\mathcal{X}} = 0
\end{equation}
where $j=1,...,N$. Finally, the system dynamics in convex formulation for the overall formation is represented as:
\begin{equation}\label{eq:FF_convex_sys_dyn_final}
\hat{\mathbf{\mathcal{A}}}_{sd} \, \hat{\mathcal{X}} = 0
\end{equation}

\paragraph{Objective function} For the cost function, we define a matrix $\mathbf{\mathcal{H}}$ to extract, from the state vector of each $j$-th satellite, the control terms $\mathbf{u}_j^k$: 
\begin{equation}\label{eq:convex_obj_fun}
\begin{array}{rl}
\mathcal{H}_j\,\hat{\mathbf{x}}_j& = \left[\mathbf{0}_{6K}, \quad \mathbf{I}_{3(K-1)}\right]\,\hat{\mathbf{x}}_j \\ 
& = \left[\mathbf{0}_{1\times 6K}, \quad \mathbf{u}_j^1, \quad \cdots, \quad \mathbf{u}_j^k, \quad \cdots, \quad \mathbf{u}_j^{K-1}\right]^T
\end{array} 	  
\end{equation}
where $j=1,...,N$. Finally, for the overall formation, the objective function including the contribution of each satellite becomes: 
\begin{equation}\label{eq:convex_obj_func_final}
J = \left\|(\hat{\mathcal{H}}\,\hat{\mathcal{X}})\,\Delta t \right\|_1
\end{equation}
where $\hat{\mathcal{H}} = \left[\cdots,\quad \mathcal{H}_j, \quad \cdots\right]^T$ for every $j=1,...,N$. Note that the expression in Eq.~(\ref{eq:convex_obj_func_final}) is equivalent to the objective function defined for the classical control problem in Eq.~(\ref{eq:performance_index}). In fact, it correspond to the sum of the norm-1 at each time instant $k$ of the control effort $\mathbf{u}_j^k$. The value from the matrix multiplication is multiplied by the discretised time interval $\Delta t$, to recover the cost of the manoeuvre in the overall time interval $\Delta T$.

\paragraph{Initial and final conditions} The same procedure used for the system dynamics and the objective function is used to write the initial and final conditions in terms of the full state column vector $\hat{\mathcal{X}}$. The final result of the described procedure is the following relation:
\begin{equation}\label{eq:in_fin_condition_convex}
\left\{\begin{array}{l}
\hat{\mathcal{A}}_{IC} \, \hat{\mathcal{X}}=\mathbf{X}_0	\\
\hat{\mathcal{A}}_{FC} \, \hat{\mathcal{X}}=\mathbf{X}_f
\end{array}
\right.
\end{equation}
where the matrices $\hat{\mathcal{A}}_{IC}$ and $\hat{\mathcal{A}}_{FC}$ are defined as:
\begin{equation}\label{eq:A_IC_convex}
\hat{\mathcal{A}}_{IC} = \left[\begin{array}{ccc}
\cdots	& \cdots & \cdots \\
\mathbf{0}_{M\times M(j-1)}	& \mathcal{A}_{IC} & \mathbf{0}_{M\times M(N-j)} \\
\cdots	& \cdots & \cdots
\end{array}	
\right]
\end{equation}
\begin{equation}\label{eq:A_FC_convex}
\hat{\mathcal{A}}_{FC} = \left[\begin{array}{ccc}
\cdots	& \cdots & \cdots  \\
\mathbf{0}_{M\times M(j-1)}	& \mathcal{A}_{FC} & \mathbf{0}_{M\times M(N-j)} \\
\cdots	& \cdots & \cdots
\end{array}	
\right]
\end{equation}
For each row defined by $j=1,...,N$,  where $\mathcal{A}_{IC}$ and $\mathcal{A}_{FC}$ are $(M\times M)$ matrices with only not-null components the $\mathcal{A}_{IC}^{(1:6,1:6)} = \mathbf{I}_6$ and $\mathcal{A}_{FC}^{(6K-5:6K,6K-5:6K)} = \mathbf{I}_6$ for each $j$-th satellite. While the $\mathbf{X}_0$ and $\mathbf{X}_f$ terms are $(6K\,N)$ column vectors for the initial/final conditions of the overall formation, defined as:
\begin{equation}\label{eq:IC_vectors}
\mathbf{X}_0 = \left[\mathbf{x}_{0,1} \quad \cdots \quad \mathbf{x}_{0,j}, \quad \cdots \quad  \mathbf{x}_{0,N} \right]^T
\end{equation} 
\begin{equation}\label{eq:FC_vectors}
\mathbf{X}_f = \left[\cdots \quad \mathbf{x}_{f,1} \quad \cdots \quad \mathbf{x}_{f,j}, \quad \cdots \quad  \mathbf{x}_{f,N} \right]^T
\end{equation}
\paragraph{Thrust limitation} For the thrust limitation, the relation is defined for multiple thrusters, i.e. each satellite $j$ could provide a thrust in a generic direction along the relative RTN frame. The thrusters are considered oriented in the body axis frame $(x_b,\,y_b,\,z_b)$, hence, depending on the attitude, it is possible to recover the components in the relative RTN frame with the relation in Eq.~(\ref{eq:RTN_to_body}). The maximum thrust given by the on-board engine poses a limit in both positive and negative directions of the firings. The relation in Eq.~(\ref{eq:thrust_lim_discrete}) is manipulated in the matrix form in terms of the full state column vector as follows:
\begin{equation}\label{eq:thrust_convex}
\hat{\mathcal{A}}_{th} \, \hat{\mathcal{X}} \leq \mathbf{a}_{max} \, \hat{\mathcal{B}}_{th}
\end{equation}
where the matrix  $\hat{\mathcal{A}}_{th}$ is defined for the multiple thruster case to extract from the full state column vector the control components $u_j^k$ for each $j=1,...,N$ and for each $k=1,...,K$.
\begin{equation}\label{eq:matrix_control_convex}
\hat{\mathcal{A}}_{th} = \left[\begin{array}{ccc}
\cdots & \cdots & \cdots \\
\mathbf{0}_{6(K-1)\times M(j-1)}& \mathcal{A}_{th} & \mathbf{0}_{6(K-1)\times M(N-j)} \\
\cdots & \cdots & \cdots 
\end{array}
\right]
\end{equation}
where $\mathcal{A}_{th} = \left[\tilde{\mathcal{A}}_{th}; -\tilde{\mathcal{A}}_{th} \right]$, and for each satellite $j$, the matrix to extract the control component from the state vector is $\tilde{\mathcal{A}}_{th}(:, 6K+1:M) =\mathbf{I}_{3(K-1)}$. Finally the column vector $\hat{\mathcal{B}}_{th}$ is defined depending on the thruster configuration:
\begin{equation}\label{eq:thrust_config}
\begin{array}{ll}
\mathrm{In\,RTN:}	& \hat{\mathcal{B}}_{th} = \mathbf{I}_{6N(K-1)\times 1}\\
\mathrm{In\,TN\,only:}	& \hat{\mathcal{B}}_{th} = \left[0,\,1,\,1,\,\cdots\,0,\,1,\,1\right]^T
\end{array}
\end{equation}

\paragraph{Inter-satellite collision avoidance constraint} The collision avoidance constraint is the most important one, since it provides a collision-free zone for the optimal manoeuvre. As defined in Eq.~(\ref{eq:coll_avoidance_convex_final}), the relation could be converted in matrix form with the full state column vector $\hat{\mathcal{X}}$ for each pair of satellites $j$ and $i$ with $j=1,...,N-1$ and $i>j$ as: 

\begin{equation}\label{eq:CA_convex_form}
\hat{\mathcal{B}}_{CA} \left(\left(\hat{\mathcal{A}}_{CA}\,\bar{\mathcal{X}}\right)^T\cdot \left(\hat{\mathcal{A}}_{CA}\,\hat{\mathcal{X}}\right)\right)\geq d_{thr}^2 \,\hat{\mathcal{C}}_{CA}
\end{equation}
where $\bar{\mathcal{X}}$ is the matrix form of the initial guess of the optimal trajectory $\bar{\mathbf{x}}_j[k]$. The matrix $\hat{\mathcal{A}}_{CA}$ is defined to extract the term $ \mathbf{x}_j[k] - \mathbf{x}_i[k]$ from the full state vector $\hat{\mathcal{X}}$ and from the initial guess $\bar{\mathcal{X}}$ at each time step $k$. The generic formulation of $\hat{\mathcal{A}}_{CA}$ for the collision avoidance constraint of satellite $i$ and $j$ is the following:
\begin{equation}\label{eq:A_ca_matrix_convex}
\hat{\mathcal{A}}_{CA} = \left[\begin{array}{c}
\vdots\\ 
\mathcal{A}_{CA}^{i,j}[k] \\ 
\vdots
\end{array} \right]
\end{equation}
\begin{equation}\label{eq:A_ca_k_matrix_convex}
\mathcal{A}_{CA}^{i,j}[k] = \left[\begin{array}{ccccc}
\mathbf{0}_{3,6(k-1)}& \mathbf{I}_{3} & \mathbf{0}_{3,3(3K-2)} & -\mathbf{I}_{3} & \mathbf{0}_{3,3(6K-2k-1)}
\end{array} \right]
\end{equation}
for $k=1,...,K$ and $j=1,...,N-1$, $i>j$. The matrix $\hat{\mathcal{B}}_{CA}$ is introduced to extract the quadratic form of the inter-satellite distance between satellite $j$ and satellite $i$.
Similarly, the matrix $\hat{\mathcal{C}}_{CA}$ is selected to represent in quadratic form the component $\mathbf{C}\left(\bar{\mathbf{x}}_j[k] - \bar{\mathbf{x}}_i[k]\right)$ of Eq.~(\ref{eq:CA_discrete}) at each time instant $k$ from the initial state $\bar{\mathcal{X}}$. The quadratic form of the collision avoidance constraint grants the convexity of the formulation, and the closer the initial guess is to the actual optimal trajectory, the more easily the optimal control problem will converge to the solution.

\subsection{Disciplined convex programming}\label{sec:disciplined_cvx_prog}
The disciplined convex programming is exploited in this article to solve the convex optimal problem defined by Eq.~(\ref{eq:convex_obj_func_final}) subject to the constraints in  Eqs.~(\ref{eq:convex_sys_dyn}), (\ref{eq:in_fin_condition_convex}), (\ref{eq:thrust_convex}), and (\ref{eq:CA_convex_form}). Both the objective and the inequality constraints are expressed in convex formulation, while the equality constraints are affine. 

There exist different software for the resolution of disciplined convex problems. An example is the \textit{SeDuMi} software, which can be employed to solve a problem involving linear and quadratic equations and inequalities, developed by \cite{sturm1999using}. A second example is the semidefinite program solver \textit{SDPT3}, an infeasible path-following algorithm for semidefinite-quadratic-linear programming developed by \cite{toh1999sdpt3}. Finally, a third similar approach is the \textit{GuRoBi} for linear and non-linear mathematical optimisation problems \citep{gurobi}. We take advantage of the sparse properties of the matrices defined in the convex problem, for a more computationally efficient resolution. Moreover, we use the \textit{CVX} Matlab\textsuperscript{\circledR} based software from \cite{grant2008graph} and  \cite{grant2013cvx}, which allows to solve a convex problem in a simple formulation, with the possibility to select either the \textit{SDPT3},  \textit{SeDuMi} or  \textit{GuRoBi} solvers. The input specifications of the \textit{CVX} software are shown in Algorithm \ref{algorithm_1}.

\begin{algorithm}[h]
	\SetAlgoLined
	\KwData{Initial and final state of each satellite $X_0$ and $X_f$\;
		\hspace{0.9cm} Maximum acceleration from the thrusters $\mathbf{a}_{max}$\;
		\hspace{1cm}Minimum inter-satellite distance $d_{thr}$; }
	\textbf{Initialisation}\;
	$\quad n = N \cdot (6K + 3 (K-1))$\;
	$\quad \Delta t = t^{k+1} - t^k$\;
	$\mathbf{cvx\_}$\Begin{
		cvx$\_$solver \hspace{0.8cm} \textit{sdpt3} (or \textit{sedumi} or \textit{gurobi})\; 
		cvx$\_$precision \hspace{0.3cm} \textit{best}\;
		variable \hspace{1.2cm} \textit{X(n)}\;
		\textit{minimise}$\left( \mathrm{norm}\left(\left(\hat{H}\,X\right)\Delta t,1\right) \right)$\;
		\textit{subject to}\;
		$\quad \hat{A}_{sd} \, X = 0$\;
		$\quad \hat{A}_{IC} \, X=X_0$\; 
		$\quad \hat{A}_{FC} \, X=X_f$\;
		$\quad \hat{A}_{th} \, X \leq \mathbf{a}_{max} \, \hat{B}_{th}$\;
		$\quad \hat{B}_{CA} \left[\left(\hat{A}_{CA}\,\bar{X}\right)\cdot \left(\hat{A}_{CA}\,X\right)\right]\geq d_{thr}^2 \,\hat{C}_{CA}$\; 
	}
	\caption{Convex optimal problem for formation reconfiguration in \textit{CVX} format}
	\label{algorithm_1}
\end{algorithm}

The \textit{SeDuMi} solver is the default solver for \textit{CVX} software, and it is typically used for most problems. Nevertheless, for some situations, the \textit{SDPT3} or \textit{GuRoBi} works better and with higher reliability \citep{grant2013cvx}. For our optimal problem, the \textit{SDPT3} was preferred for a more stable solution, when the collision-avoidance constraint is employed. To guarantee a more accurate and reliable solution, the tolerance level of the solver is set at to $[\epsilon^{1/2},\epsilon^{1/2},\epsilon^{1/4}]$, where $\epsilon = 2.22 \times 10^{-16}$ is the machine precision. Note that the dimension of the problem should be defined before the call to the \textit{CVX} solver. For our problem, the dimension of the full state variable depends on the number of satellites $N$ and on the number of time step $K$. 
A consideration about the time step should be done, before passing to the study case results in Section \ref{sec:study_Case}. The time step is selected as a fraction of the orbital period of the reference orbit of the formation. It is important to consider a fraction of the orbital period to guarantee the convergence of the optimal problem. In particular, for an accurate solution and to guarantee the convergence, the discretisation for a manoeuvre in one orbital period should not be higher than $2°/n_c$, where $n_c$ is the mean motion of the reference orbit \citep{sarno2020guidance}.

		\subsection{Algorithm performances evaluation}\label{sec:performances_cvx_prog}
\begin{figure}[htb]
	\centering
	\includegraphics[width=0.35\textwidth]{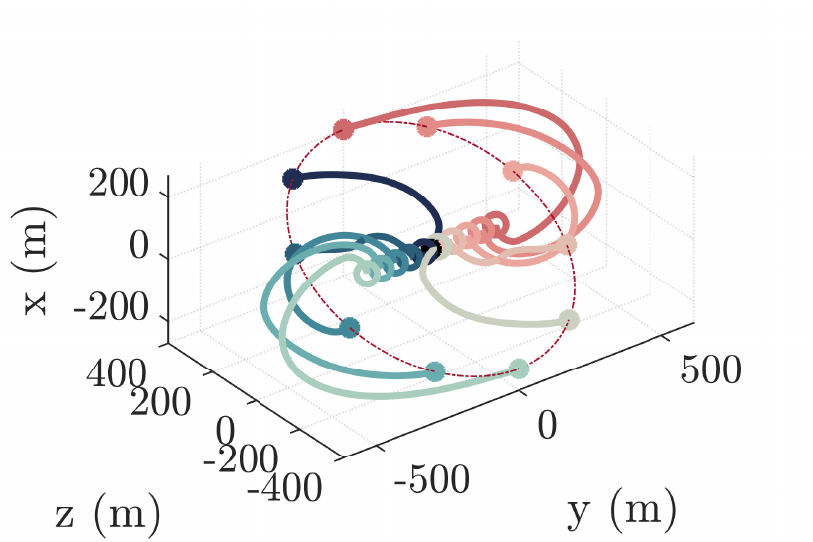}
	\caption{Coplanar to PCO reconfiguration for a 10-satellite formation.}
	\label{fig:Swarm_reconfig}
\end{figure}

		The performances of the selected methodology based on disciplined convex programming are tested against a progressive number of satellites (from 2 to 12), to evaluate the computational time. The different solvers available for the \textit{CVX} software, as described in Section~\ref{sec:disciplined_cvx_prog} are tested for performance evaluation. The test case presented in this section simulates a classic coplanar to Projected Circular Orbit (PCO) \citep{alfriend2009spacecraft}, considering the simulation parameters reported in Table~\ref{tab:Swarm_Properties}. 
		\begin{table*}[b]
			\centering
			\caption{Parameters for simulation setting.}
			\label{tab:Swarm_Properties}
			\begin{tabular}{|ll|l|}
				\hline
				\textbf{Properties}  &  & \textbf{Value} \\
				\hline
				Manoeuvring time &  [s] & $3/4\,P$   \\
				\hline
				Discretisation step & [sec] & 25 \\
				\hline
				Minimum inter-satellite distance & [m] & 20 \\
				\hline
				Satellites Mass & [kg] & 50 \\
				\hline
				Maximum Thrust & [mN] & 10   \\
				\hline
			\end{tabular}
		\end{table*}		
		The reference orbit is an SSO with the Keplerian elements equal to $\{7.416e6\,\mathrm{m},\,0,\,98.5^\circ,\,0^\circ,\,0^\circ\}$.
		The satellites are initially placed in a coplanar formation configuration along the transversal direction, with an initial inter-satellite distance of 50 meters. The final condition for each satellite $j$ is selected as a PCO using the procedure described in \cite{alfriend2009spacecraft}, based on magnitude-phase form of the relative motion:
		\begin{equation}
		\label{eq:HCW_magn_phase}
		\left\{\begin{array}{l}
		\delta a = 0 \\
		\delta \lambda = \Big(\frac{\rho}{a_c}\Big) \frac{\sin \alpha_j}{\tan i_c}\\
		\delta e_x = - \Big(\frac{\rho}{2 a_c}\Big) \sin \alpha_j \\
		\delta e_y = - \Big(\frac{\rho}{2 a_c}\Big) \cos \alpha_j \\
		\delta i_x = \Big(\frac{\rho}{a_c}\Big) \cos \alpha_j \\
		\delta i_y = - \Big(\frac{\rho}{a_c}\Big) \sin \alpha_j 
		\end{array} \right.
		\end{equation}	
		Where $\rho = 200 $ m is the radius of the PCO in the transversal-normal plane, while $\alpha_j$ is the phase angle for each satellite, selected to get an equally distributed formation along the final orbit.	
		An example of the reconfiguration trajectory for the 10-satellites formation case is shown in Figure~\ref{fig:Swarm_reconfig}. The solution is obtained imposing a first guess of the trajectory without collision avoidance constraint and then refining the solution with a second iteration of the optimisation. This procedure guarantees the correct inclusion of the collision avoidance constraint in the simulation.

\begin{figure}[htb]
	\centering
	\begin{subfigure}[b]{0.30\textwidth}
		\centering
		\includegraphics[width=\textwidth]{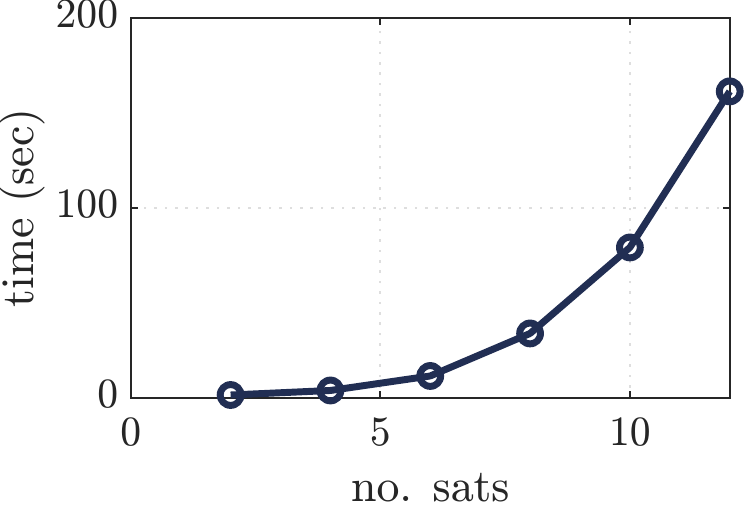}
		\caption{Computational time for setting up the simulation.}
		\label{fig:Matrix_perform}
	\end{subfigure}
	\begin{subfigure}[b]{0.30\textwidth}
		\centering
		\includegraphics[width=\textwidth]{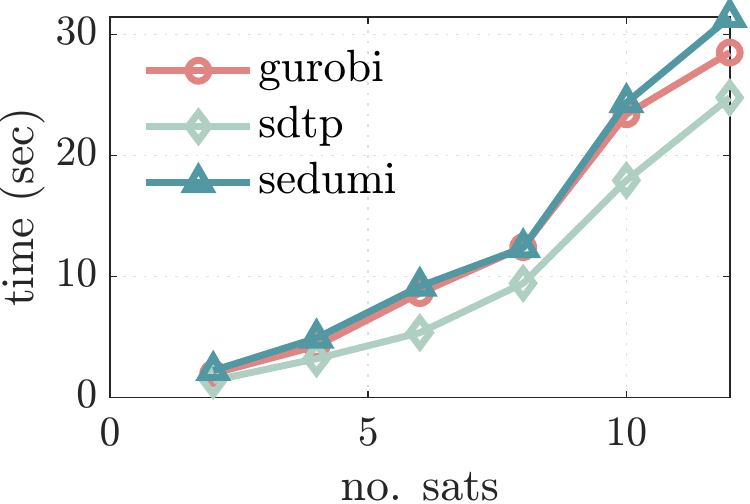}
		\caption{Computational time for CVX solver.}
		\label{fig:CVX_perform}
	\end{subfigure}	
	\caption{Algorithm performance evaluation with 2 to 12 satellites formation.}
	\label{fig:Swarm_perform}
\end{figure}

		Figure~\ref{fig:Swarm_perform} show the performances of the algorithm against the number of satellites. 
		Figure~\ref{fig:CVX_perform}	represents the computational time to set up the matrices for the convex problem, which is required to initialised the \textit{CVX} software. The behaviour is exponential with the increasing number of satellites, varying from a minimum of 2 seconds to a maximum of about 3 minutes for the 12-satellites case. This time is required only for the initialization of the convex problem and does not affect the \textit{CVX} solver time.
		Figure~\ref{fig:Matrix_perform} shows the computational time required by the \textit{CVX} solver to provide the optimal solution, considering the three possible solvers of \textit{CVX}, \textit{GuRoBi}, \textit{STDP3} and \textit{SeDuMi}. In this case, the behaviour scales approximately linear with the number of satellites, with a maximum of about 30 seconds for the 12 satellites case. For the coplanar to PCO reconfiguration, the \textit{STDP3} provides slightly better performances with a higher number of satellites.

\section{Application to the study case: FFLAS}\label{sec:study_Case}

The proposed methodology developed in Section \ref{sec:methodology} is applied to a formation flying mission concept for remote sensing and Earth observation. The generic mission scenario was described in Section \ref{sec:ProblemDef}, where the importance of using multiple satellites as distributed nodes is described. In this work, we consider a mission scenario made by three satellites, embarking each a hexagonal L-band antenna array \citep{Zurita2013,Neira2020}. The geometry trade-off analyses and the mission concept were described by \cite{scala2020formation}. The concept of the mission takes advantage of the basic parameter for the L-band payload, resulting from the lessons learnt of the SMOS mission \citep{kerr2016overview,Neira2020}. The three-satellite formation flies on a Sun-Synchronous Orbit (SSO) at a nominal altitude of 770 km, with a Local Time of the Ascending Node (LTAN) of 6:00 a.m. The L-band aperture synthesis payload is selected as a hexagonal array, of about 7 m in diameter, a slightly smaller size than the SMOS one. The centres of each satellite in the formation are placed at the vertices of an equilateral triangle of about 13 m sides. Each satellite weight about 1300 kg of dry mass and has four low-thrust engines onboard, considering as baseline the QuinetiQ T5 \citep{randall2017qinetiq}. 
Two main modes have been identified for the nominal operation in orbit. The Earth pointing mode for the L-band aperture synthesis imaging the Earth's surface, and the cold sky pointing mode, for the calibration of the interferometer. This work presents the possible strategies for the transition among the two nominal modes of the mission, together with a safe-mode transition design. This is of primary importance in FFLAS, for a safe formation definition in case of non-nominal situations arises. Finally, it is also presented the satellite-to-satellite collision avoidance manoeuvre in case of a failure of the main engines of one satellite in the formation.

\begin{figure}[!htb]
	\centering
	\begin{subfigure}[b]{0.28\textwidth}
		\centering
		\includegraphics[width=\textwidth]{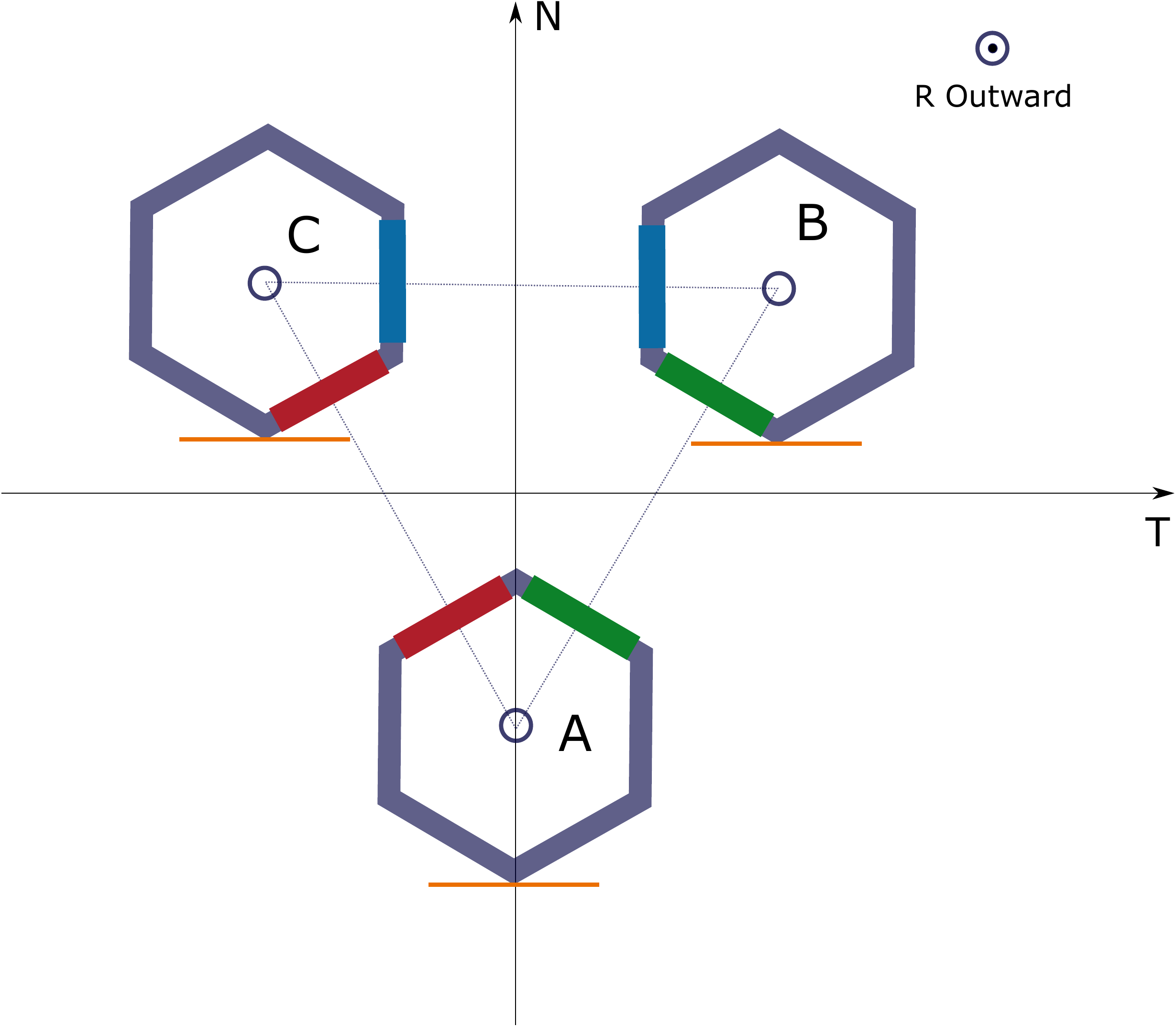}
		\caption{EPM configuration.}
		\label{fig:EPM_config}
	\end{subfigure}
	\hspace{1em}
	\begin{subfigure}[b]{0.28\textwidth}
		\centering
		\includegraphics[width=\textwidth]{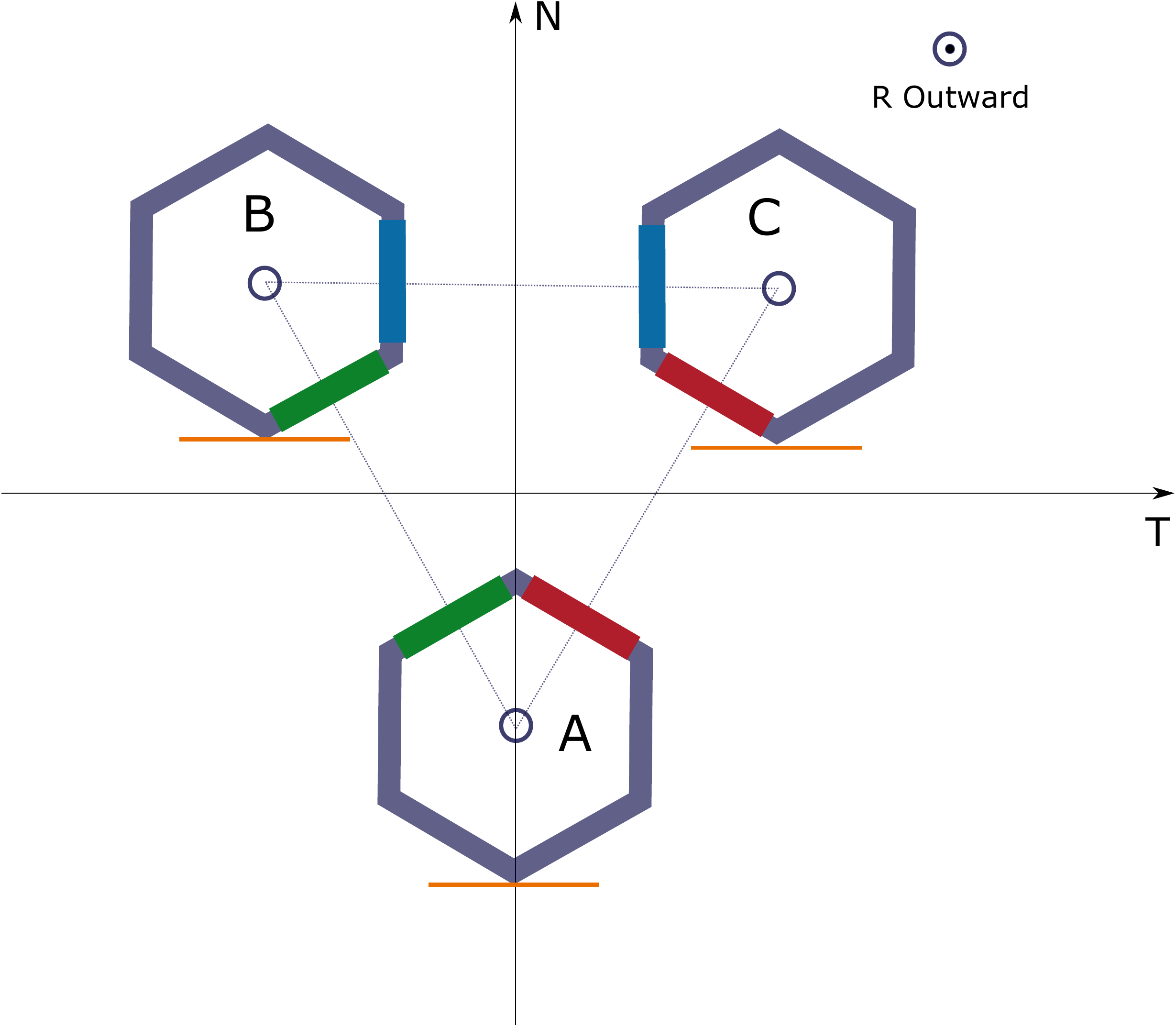}
		\caption{CSPM configuration.}
		\label{fig:CSPM_config}
	\end{subfigure}
	\caption{EPM and CSPM configuration for the three-satellite formation.}
	\label{fig:FFLAS_config}
\end{figure}

\paragraph{Earth Pointing Mode (EPM)}
The Earth Pointing Mode is the nominal operational mode of FFLAS. In this mode, the satellites in the formation should maintain the normal to the payload aligned with the radial reference axis, in the direction of the centre of the Earth. As a consequence, the nominal attitude of each satellite in the body frame is the following (Section \ref{sec:body_frame}):
\begin{equation}\label{eq:EPM_attitude}
\left\{\begin{array}{c}
x_b\\ 
y_b\\ 
z_b
\end{array} \right\} = \left[\begin{array}{ccc}
-1& 0 & 0 \\ 
0& 1 & 0 \\ 
0& 0 & -1
\end{array} \right] \left\{\begin{array}{c}
x\\ 
y\\ 
z
\end{array} \right\}
\end{equation}
where the x-axis $x_b$ of the spacecraft points in the geocentric nadir direction for observation purposes, and the y-axis $y_b$ of the spacecraft is in the orbit vector direction. The EPM formation geometry for the test case mission study FFLAS is shown in Fig.~\ref{fig:EPM_config}. The solar panels (the orange lines in Fig.~\ref{fig:EPM_config}) are in the Sun direction, to maximise the power production, and the attitude of the formation is selected to allow the inter-satellite communication link, to exchange continuously the raw data necessary for the interferometric. The inter-satellite link antennas are shown in green, blue and red in Fig.~\ref{fig:EPM_config}, to represent the optical link geometry.


\paragraph{Cold Sky Pointing Mode (CSPM)}
The Cold Sky Pointing Mode is the calibration mode for the aperture synthesis radiometer payload. The calibration of the payload is essential for a good quality interferometric imaging and it is typically required at least once per month. During this mode, the z‐axis $z_b$ of each satellite should be in the cold sky inertial direction. The transition to the CSPM attitude is designed with the optimal manoeuvre procedure, presented in Section \ref{sec:methodology}. The CSPM formation geometry for FFLAS is shown in Fig.~\ref{fig:CSPM_config}. With respect to the EPM configuration, it is shown how the satellites B and C switch their position to provide a cold sky pointing attitude, but maintaining the correct optical link among them. Moreover, to respect the cold sky pointing direction, an attitude slew manoeuvre of 180° is needed for each satellite, to move the pointing direction of the payload towards the cold sky. The attitude of the satellites in the CSPM is described by the following relation:
\begin{equation}\label{eq:CSPM_attitude}
\left\{\begin{array}{c}
x_b\\ 
y_b\\ 
z_b
\end{array} \right\} = \left[\begin{array}{ccc}
1& 0 & 0 \\ 
0& 1 & 0 \\ 
0& 0 & 1
\end{array} \right] \left\{\begin{array}{c}
x\\ 
y\\ 
z
\end{array} \right\}
\end{equation}
Where the body frame is aligned with the radial-transversal-normal frame, and the interferometric radiometer points to the cold sky direction.

\paragraph{Thruster configuration} The analyses presented here consider the thrusters placed in the Transversal-Normal plane (TN) only. The motivation behind this selection is based on the need of reducing the mass budget of the satellites. Moreover, this is needed for satellite internal configuration purposes. Thus, no thrust in the radial direction could be provided, and the acceleration is given only in the $y_b$ and $z_b$ axis of the satellite. As a consequence, in the optimal control problem, the thrust limitation is given by the second relation in Eq.~(\ref{eq:thrust_config}). The maximum thrust level that the thrusters could provide is set equal to 25 mN, considering as a baseline for the propulsive system, the QinetiQ T5 engine \citep{randall2017qinetiq}. 

\subsection{Optimal manoeuvre strategy from Earth pointing to cold sky pointing mode}
The transition between the EPM and the CSPM is designed with a fuel-optimal control problem, considering the following requirements:
\begin{itemize}
	\item The attitude of the satellites should be compliant with the Sun direction, to ensure enough power generation,
	\item The transition should be performed in less than one orbital period and a two-axis thruster configuration is considered,
\end{itemize}
The inter-satellite collision risk is managed by implementing two cases for the minimum distance among the satellites: 10 m (Case i) and 12 m (Case ii). This means that at each time instant, the distance among the satellites should not violate such conditions, to ensure a safe transition to the calibration mode.
The optimal problem was initialised considering the conditions in Table~\ref{tab:EPM2CSPM_Properties}. The time for propagation is provided in terms of orbital periods $P$, and the time step represents the discretisation step along the orbit. The total time of the transition is selected equal to $3/4$ of the orbital period to deal with the constraint on the maximum available thrust and the minimum allowable inter-satellite distance.
The \textit{CVX} problem was initialised with the semi-definite quadratic-linear programming SDP3, accordingly to the convex optimisation problem described in \ref{sec:disciplined_cvx_prog}. On a Windows computer with Intel(R) Core(TM) i7-4720HQ CPU @ 2.60GHz at 2.59 GHz and a RAM of 16.0 GB, the convex optimisation is solved in about 2.5 seconds for the case of a minimum inter-satellite distance of 10 m, and in about 5 seconds for the 12 m case. 

\begin{table*}[!htb]
	\centering
	\caption{Input conditions for the optimal manoeuvre design between earth pointing and cold sky point mode.}
	\label{tab:EPM2CSPM_Properties}
	\begin{tabular}{|ll|l|}
		\hline
		\textbf{Properties}  &  & \textbf{Value} \\
		\hline
		Manoeuvring time&  [s] & $3/4\,P$   \\
		\hline
		Discretisation step& [sec] & 25 \\
		\hline
		Minimum distance& [m] & 10 and 12 \\
		\hline
		State of Sat A& [m, m/s] &$\left[0,\quad 0,\quad -5.6,\quad 0,\quad 0,\quad 0\right]$ \\
		\hline
		Initial state - Sat B& [m, m/s] & $\left[0,\quad \,\,\,\, 6.5,\quad 5.6,\quad 0,\quad 0,\quad 0\right]$   \\
		\hline
		Initial state - Sat C& [m, m/s] & $\left[0,\quad -6.5,\quad 5.6,\quad 0,\quad 0,\quad 0\right]$   \\
		\hline
		Final state - Sat B& [m, m/s] & $\left[0,\quad -6.5,\quad 5.6,\quad 0,\quad 0 ,\quad 0\right]$   \\
		\hline
		Final state - Sat C &[m, m/s] & $\left[0,\quad \,\,\,\, 6.5,\quad 5.6,\quad 0,\quad 0,\quad 0\right]$   \\
		\hline
	\end{tabular}
\end{table*}

\begin{figure}[!htb]
	\centering
	\begin{subfigure}[b]{0.35\textwidth}
		\centering
		\includegraphics[width=\textwidth]{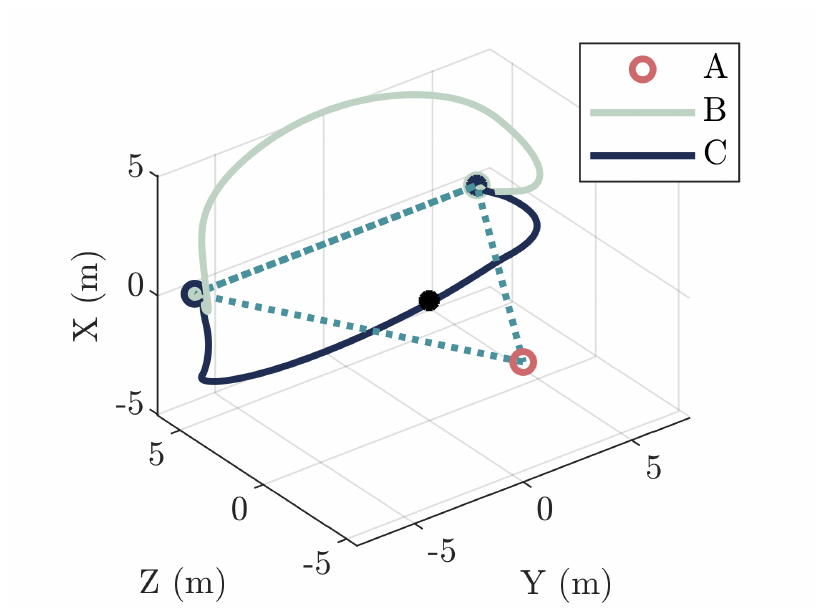}
		\caption{Optimal trajectory in RTN frame.}
		\label{fig:EPM2CSPM_10}
	\end{subfigure}
	\hspace{1.5em}
	\begin{subfigure}[b]{0.35\textwidth}
		\centering
		\includegraphics[width=\textwidth]{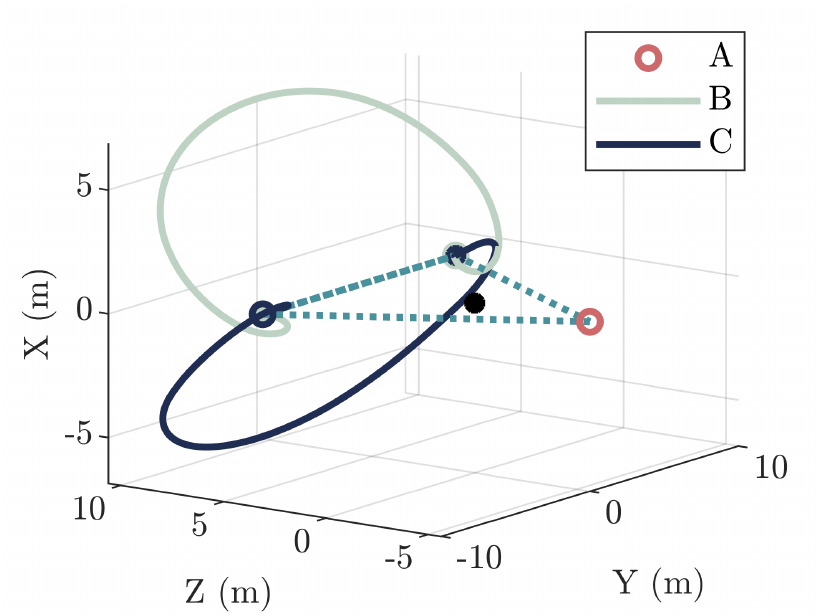}
		\caption{Optimal trajectory.}
		\label{fig:EPM2CSPM_12}
	\end{subfigure}
	\hspace{1.5em}
	\begin{subfigure}[b]{0.30\textwidth}
		\centering
		\includegraphics[width=\textwidth]{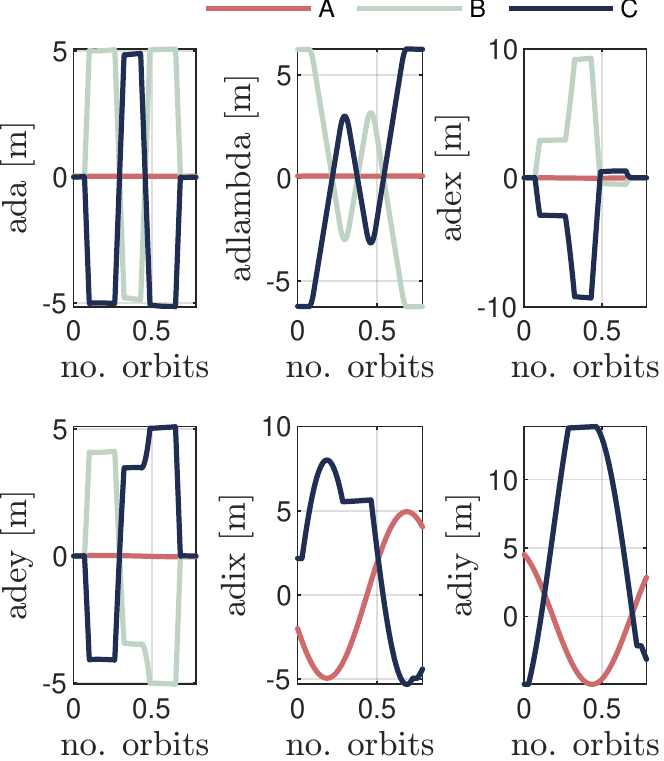}
		\caption{Time evolution of the ROEs components.}
		\label{fig:EPM2CSPM_10_ROEs}
	\end{subfigure}
	\hspace{1.5em}
	\begin{subfigure}[b]{0.30\textwidth}
		\centering
		\includegraphics[width=\textwidth]{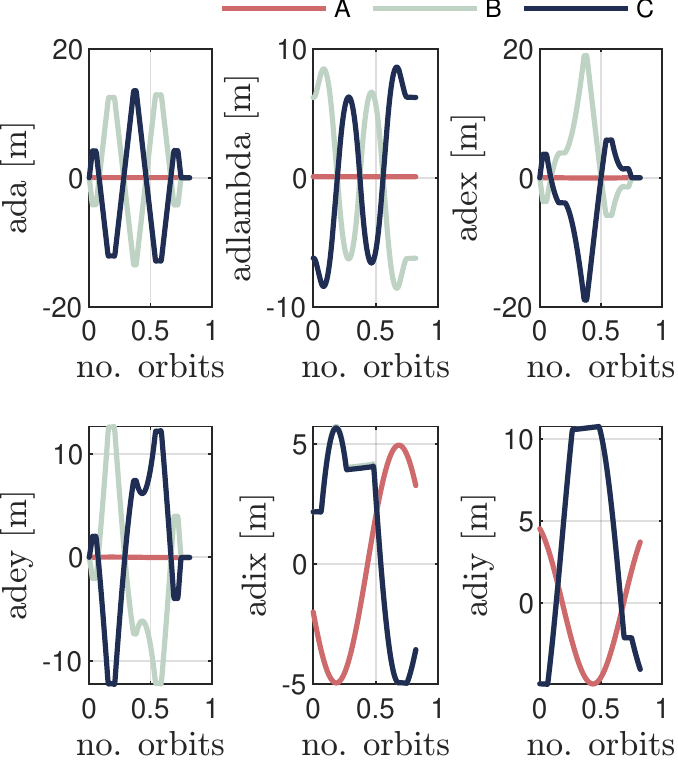}
		\caption{Time evolution of the ROEs components.}
		\label{fig:EPM2CSPM_12_ROEs}
	\end{subfigure}	
	\caption{Optimal manoeuvre for Earth pointing to cold sky pointing transition for Case i, with minimum inter-satellite threshold of 10 m (left) and 12 m (right).}
	\label{fig:EPM2CSPM_CI_Traj}
\end{figure}

\begin{figure}[!htb]
	\centering
	\begin{subfigure}[b]{0.30\textwidth}
		\centering
		\includegraphics[width=\textwidth]{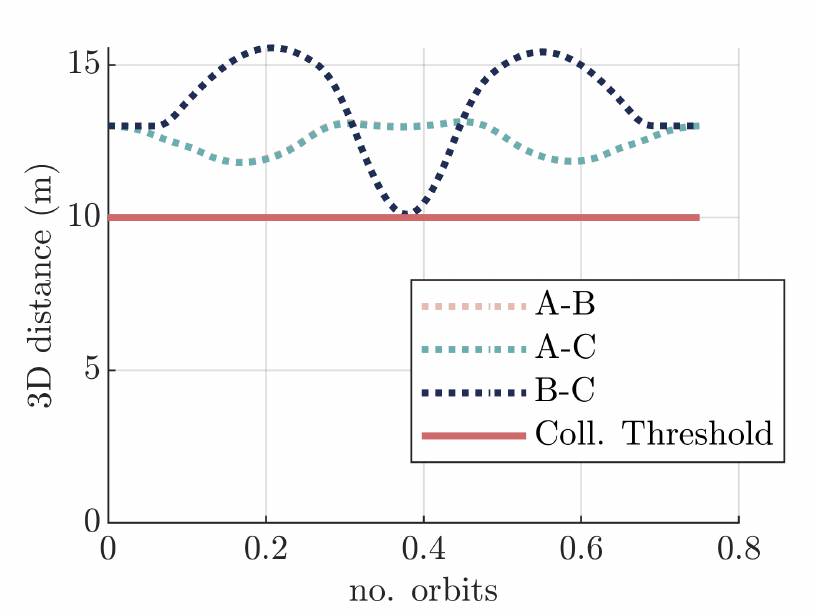}
		\caption{Inter-satellite distance among the satellite.}
		\label{fig:EPM2CSPM_10_CA}
	\end{subfigure}
	\begin{subfigure}[b]{0.30\textwidth}
		\centering
		\includegraphics[width=\textwidth]{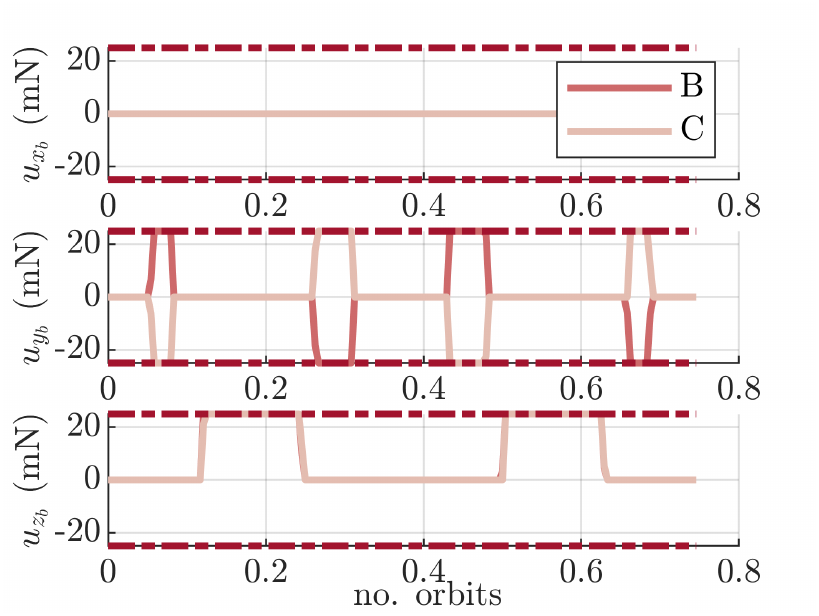}
		\caption{Control effort for satellite B and C.}
		\label{fig:EPM2CSPM_10_thrust}
	\end{subfigure}
	\begin{subfigure}[b]{0.30\textwidth}
		\centering
		\includegraphics[width=\textwidth]{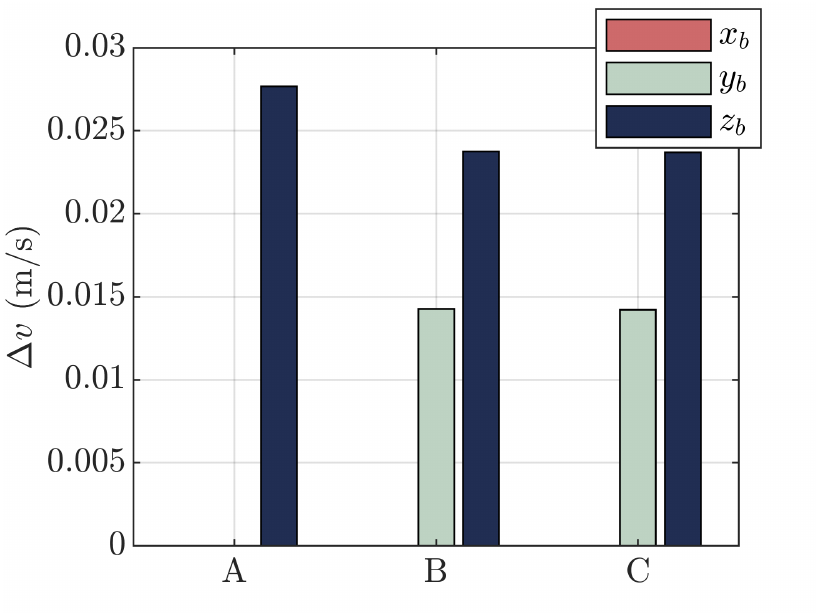}
		\caption{Delta-V budget for sat. A, B and C.}
		\label{fig:EPM2CSPM_10_DV}
	\end{subfigure}
	\begin{subfigure}[b]{0.30\textwidth}
		\centering
		\includegraphics[width=\textwidth]{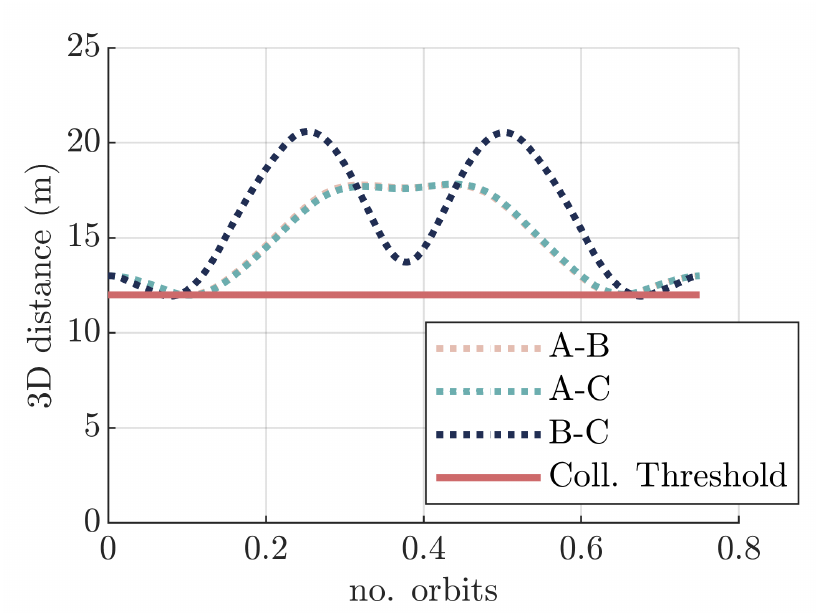}
		\caption{Inter-satellite distance among the satellite.}
		\label{fig:EPM2CSPM_12_CA}
	\end{subfigure}	
	\begin{subfigure}[b]{0.30\textwidth}
		\centering
		\includegraphics[width=\textwidth]{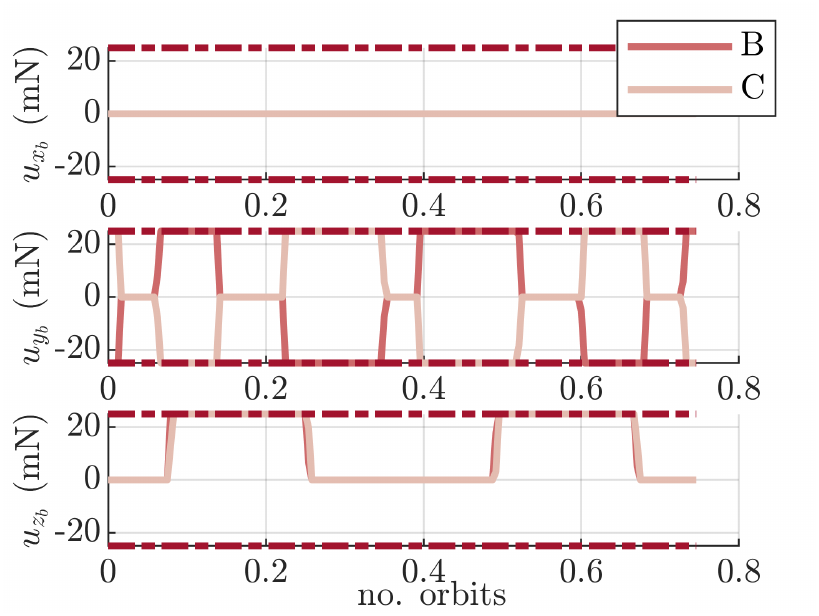}
		\caption{Control effort for satellite B and C.}
		\label{fig:EPM2CSPM_12_thrust}
	\end{subfigure}
	\begin{subfigure}[b]{0.30\textwidth}
		\centering
		\includegraphics[width=\textwidth]{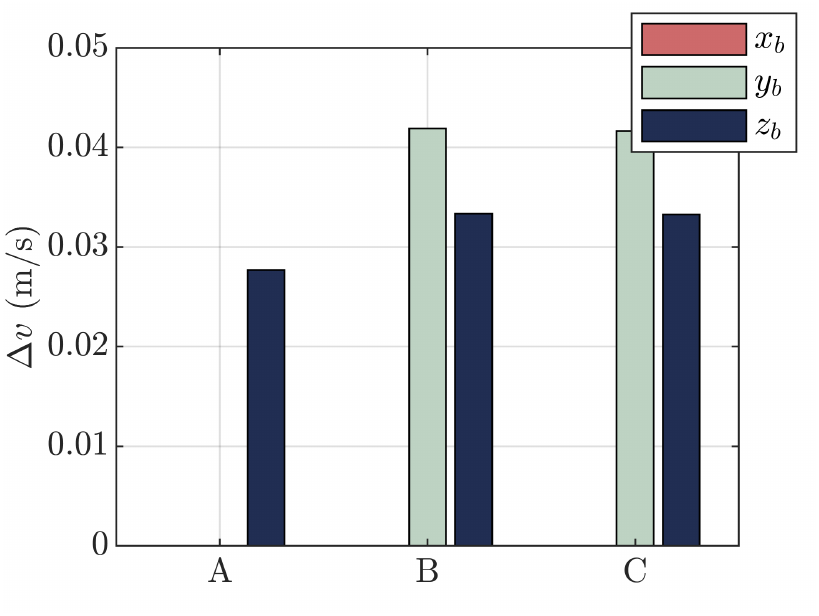}
		\caption{Delta-V budget for sat. A, B and C.}
		\label{fig:EPM2CSPM_12_DV}
	\end{subfigure}	
	\caption{Optimal manoeuvre for Earth pointing to cold sky pointing transition for Case i, with minimum inter-satellite threshold of 10 m (top) and 12 m (bottom).}
	\label{fig:EPM2CSPM_CI_thrust}
\end{figure}

\paragraph{Case i} The optimal trajectory for the manoeuvre is shown in Fig.~\ref{fig:EPM2CSPM_CI_Traj}, where Fig.~\ref{fig:EPM2CSPM_10} reports the 3-dimensional trajectory evolution in the RTN frame, while Fig.~\ref{fig:EPM2CSPM_10_ROEs} shows the time evolution of the ROEs components during the delta-v optimal manoeuvre. An important parameter to monitor is the inter-satellite distance, to control the feasibility of the mission itself, as shown in Fig.~\ref{fig:EPM2CSPM_10_CA}. This information is particularly relevant for FFLAS since the satellites fly at a close distance from each other. The standard control effort (as for the EPM) for the satellite A maintenance is required, and the control behaviour during the manoeuvre transition for satellites B and C is shown in Fig.~\ref{fig:EPM2CSPM_10_thrust}. 

Due to a quasi symmetrical behaviour in the trajectory followed by satellites B and C, it can be seen how the control effort in the normal direction ($z_b$) is coincident for both satellites. Moreover, due to the requirement of having thrusters aligned with the transversal and normal ($y_b$ and $z_b$) direction, for both satellites the control effort in the radial direction is null. The red dot lines represent the technological limitation of the onboard engine, with a maximum thrust equal to 25 mN. The delta-v budget for Case i is shown in Fig.~\ref{fig:EPM2CSPM_10_DV}, resulting in a total delta-v for the optimal transition from Earth pointing to cold sky pointing of about 3.801 cm/s. 

\paragraph{Case ii} The delta-v optimal trajectories for Case ii is shown in Figure~\ref{fig:EPM2CSPM_CI_Traj}, providing both the three-dimensional representation in RTN frame and the ROEs time evolution in Figures~\ref{fig:EPM2CSPM_12} and \ref{fig:EPM2CSPM_12_ROEs}, respectively. The trajectory is similar to Case i, except for the constrain to maintain a higher inter-satellite distance of 12 m. The standard control effort (as for the EPM) for satellite A is required, while the control behaviour for satellites B and C is shown in Fig.~\ref{fig:EPM2CSPM_12_thrust}. As for Case i, the control effort in the normal direction ($z_b$) is coincident for both satellites. Moreover, for all satellites the control effort in the radial direction is null. Overall, the control effort and the time of firings is higher than for Case i since the problem is more constrained from a collision avoidance point of view, as shown in the delta-v budget in Fig.~\ref{fig:EPM2CSPM_12_DV}. 
The total delta-v for the optimal transition increases to about 7.523 cm/s. This is due to a higher safety level in the minimum allowed inter-satellite distance. 

The control of satellite A, together with satellites B and C are needed to ensure a safe reconfiguration, controlling the inter-satellite distance continuously during the manoeuvre, for both cases. The real-time minimum inter-satellite distance computation is shown in Fig.~\ref{fig:EPM2CSPM_12_CA}, where the requirements are met through optimisation. This second case grants higher confidence in the collision avoidance risk among the satellites of the formation but requires a delta-v for satellites B and C twice the one in the first case. The trade-off must be done depending on the final delta-v budget and in the collision avoidance level to be maintained during operations. 
Note that the time evolution of the distance between satellites A and B has the same behaviour as the A-C distance. This is due to the symmetric properties of the optimal trajectory obtained in the analysis.

\subsection{Safe-Mode}\label{sec:safe_mode}

This section presents the manoeuvre transition to Safe Mode. The transfer strategy relies on the optimal delta-v trajectory design exploiting the convex optimal control problem. The followings constraints were considered:
\begin{itemize}
	\item A two-axis thruster configuration is considered in the transversal and normal axis of the body frame,
	\item The inter-satellite collision risk is managed by setting the minimum distance among the satellites of 10 m,
	\item The attitude of the satellite should remain fixed during the manoeuvre, to ensure the correct thrusting in the normal and transversal directions.
\end{itemize}
The optimal problem was initialised considering the conditions in Table~\ref{tab:EPM2SM_Properties}. The time for propagation is provided in terms of orbital periods $P$, and the time step represents the discretisation step along the orbit. The propagation time was selected to be $3/4$ of the orbital period, as before, for a trade-off among the need for a fast transition to the safe mode in case of non-nominal situations and the constraints on the maximum available thrust and the minimum allowable inter-satellite distance. The CVX problem was initialised with the semi-definite quadratic-linear programming SDP3 \citep{grant2013cvx}.
Two cases were considered for the Safe Mode formation geometries:
\begin{enumerate}
	\item Increasing the formation baseline of the aperture angle in the Radial-Normal (RN) plane,
	\item Introducing an RN separation margin for the passive safety sufficient condition.
\end{enumerate}

\begin{table*}[htb]
	\centering
	\caption{Input conditions for the optimal manoeuvre design between Earth pointing and safe mode.}
	\label{tab:EPM2SM_Properties}
	\begin{tabular}{|l|l|}
		\hline
		\textbf{Properties}    & \textbf{Value} \\
		\hline
		Manoeuvring time [s] & $3/4\,P$   \\
		\hline
		Discretisation step [sec] & 25 \\
		\hline
		Minimum inter-satellite distance [m] & 10  \\
		\hline
	\end{tabular}
\end{table*}

\paragraph{CASE I:} The formation aperture baseline was increased from 13 m (nominal side of the equilateral triangle) to 50 m and the aperture plane was maintained on the TN plane. Moreover, the relative attitude among the satellites was retained to ensure the correct inter-satellite link. Passive safety was considered by evaluating the distance during the natural formation evolution.
\begin{figure}[!htb]
	\centering
	\begin{subfigure}[b]{0.35\textwidth}
		\centering
		\includegraphics[width=\textwidth]{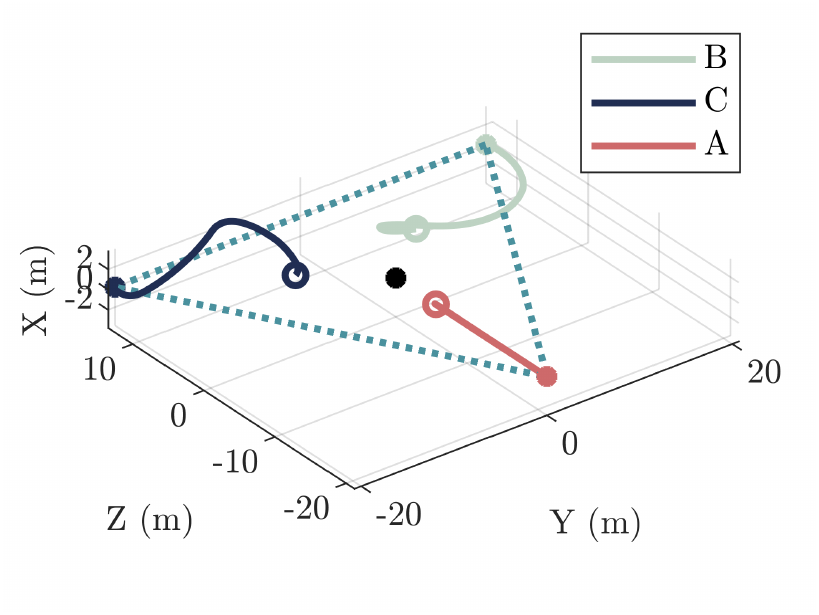}
		\caption{Optimal trajectory in RTN frame.}
		\label{fig:EPM2SM_1}
	\end{subfigure}
	\hspace{0.5cm}
	\begin{subfigure}[b]{0.30\textwidth}
		\centering
		\includegraphics[width=\textwidth]{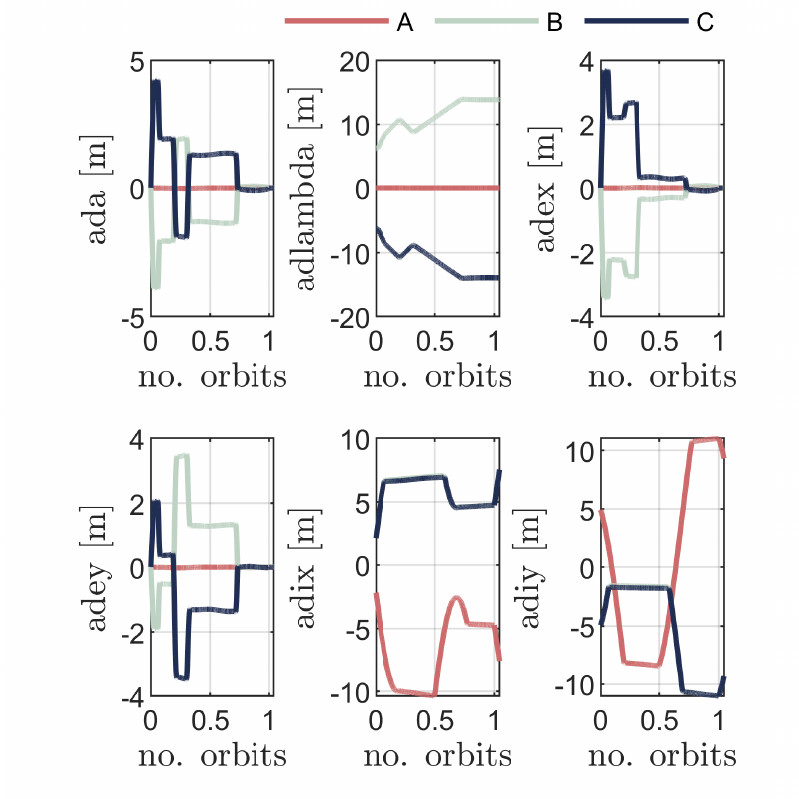}
		\caption{Time evolution of the optimal trajectory in ROEs.}
		\label{fig:EPM2SM_ROEs_1}
	\end{subfigure}
	\hspace{1.5cm}
	%
	\begin{subfigure}[b]{0.30\textwidth}
		\centering
		\includegraphics[width=\textwidth]{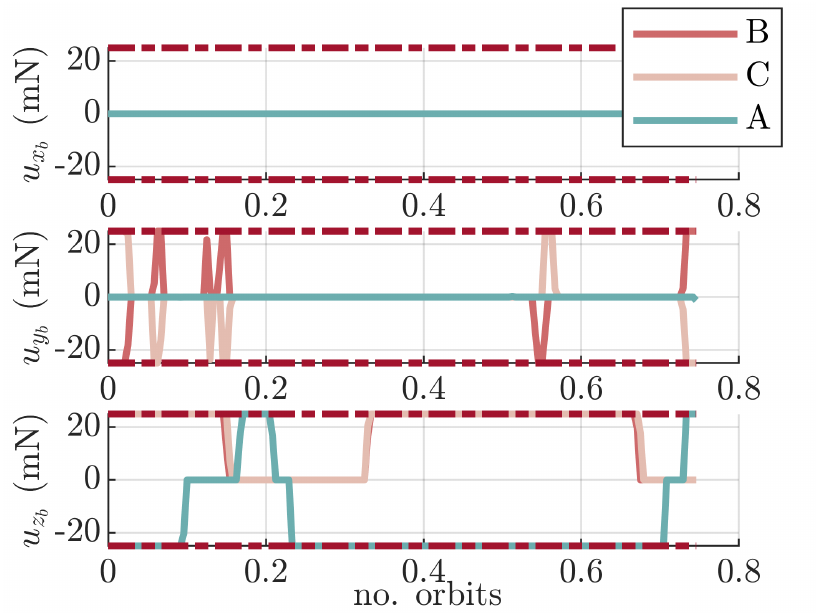}
		\caption{Control effort for satellite B and C.}
		\label{fig:EPM2SM_1_thrust}
	\end{subfigure}
	\hspace{1cm}
	\begin{subfigure}[b]{0.30\textwidth}
		\centering
		\includegraphics[width=\textwidth]{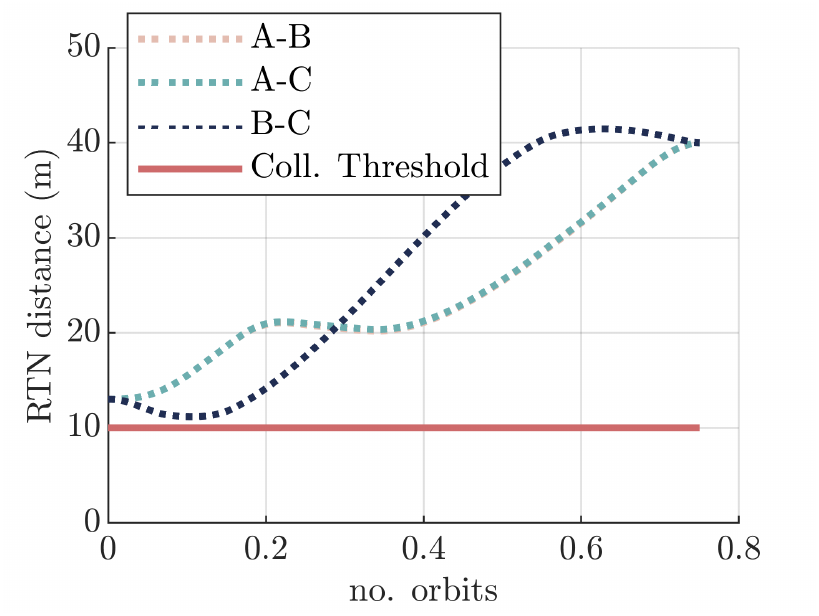}
		\caption{Inter-satellite distance among the satellites.}
		\label{fig:EPM2SM_1_CA}
	\end{subfigure}
	\hspace{1cm}
	\caption{Optimal manoeuvre for Earth pointing to safe mode transition for Case I. Minimum inter-satellite threshold set to 10 m.}
	\label{fig:EPM2SM_varius_1}
\end{figure}

\begin{figure}[!htb]
	\centering
	\hspace{1cm}
	\begin{subfigure}[b]{0.35\textwidth}
		\centering
		\includegraphics[width=\textwidth]{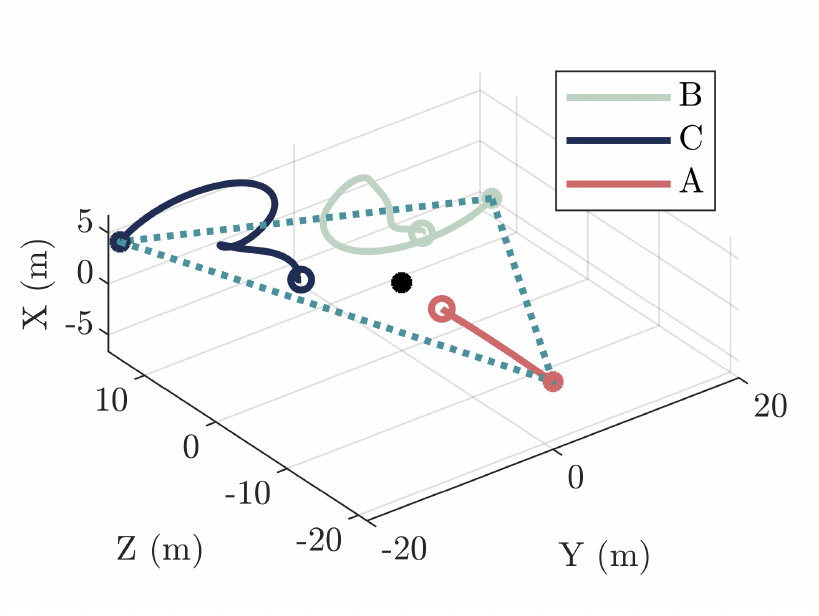}
		\caption{Optimal trajectory in RTN frame.}
		\label{fig:EPM2SM_2}
	\end{subfigure}	
	\hspace{1cm}
	\begin{subfigure}[b]{0.30\textwidth}
		\centering
		\includegraphics[width=\textwidth]{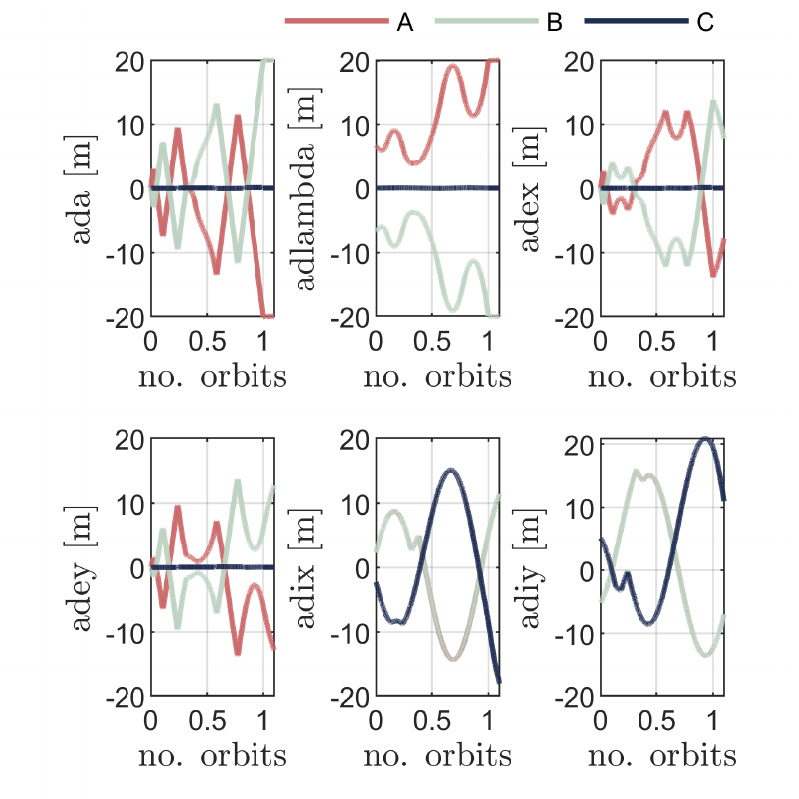}
		\caption{Time evolution of the optimal trajectory in ROEs.}
		\label{fig:EPM2SM_ROEs_2}
	\end{subfigure}
	\hspace{1cm}
	\begin{subfigure}[b]{0.30\textwidth}
		\centering
		\includegraphics[width=\textwidth]{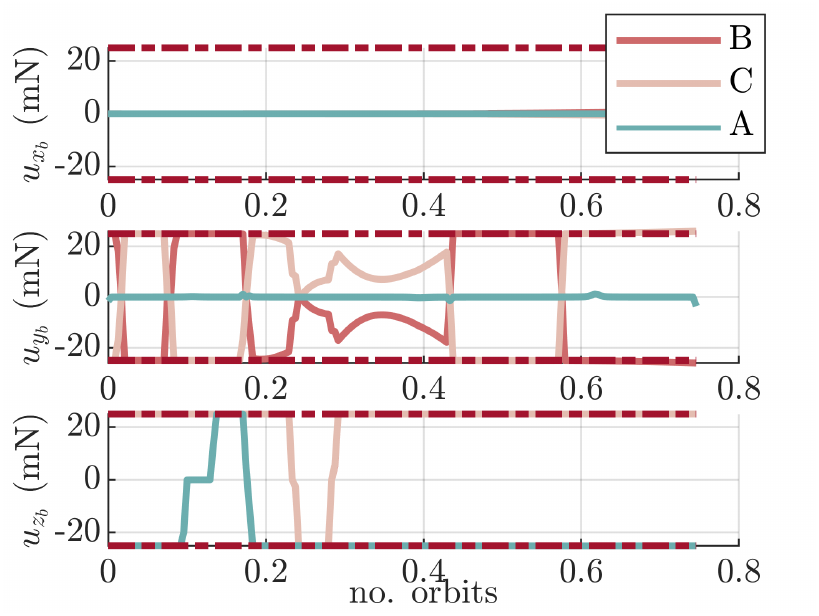}
		\caption{Control effort for satellite B and C for Case ii.}
		\label{fig:EPM2SM_2_thrust}
	\end{subfigure}	
	\hspace{1cm}
	\hspace{1cm}
	\begin{subfigure}[b]{0.30\textwidth}
		\centering
		\includegraphics[width=\textwidth]{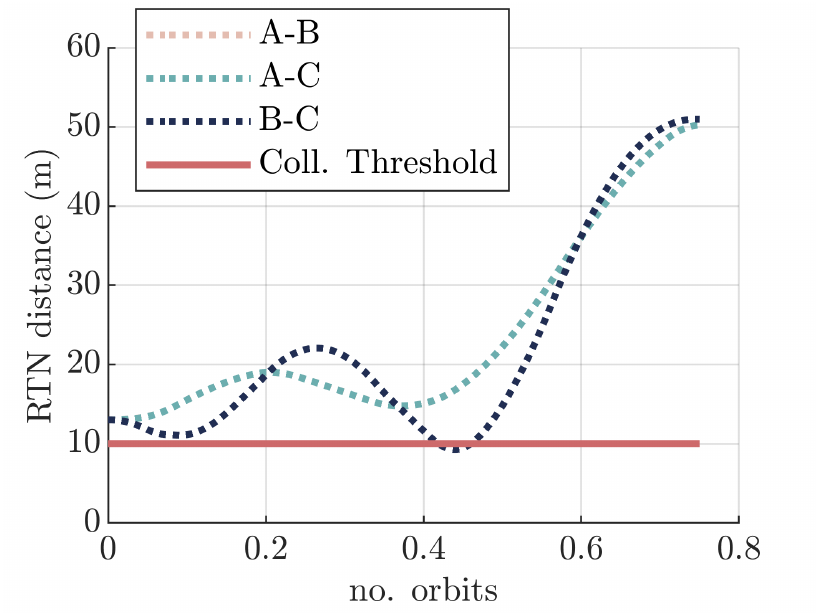}
		\caption{Inter-satellite distance among the satellites.}
		\label{fig:EPM2SM_2_CA}
	\end{subfigure}
	\caption{Optimal manoeuvre for Earth pointing to safe mode transition for Case II. Minimum inter-satellite threshold set to 10 m.}
	\label{fig:EPM2SM_varius_2}
\end{figure}

Fig.~\ref{fig:EPM2SM_1_CA} shows how the satellites remain safely far apart from one to each other, with no collision risk. The optimal trajectory for the transition from EPM to SM is shown in Fig.~\ref{fig:EPM2SM_1} and \ref{fig:EPM2SM_ROEs_1}. The trajectories for satellites B and C are quite similar to each other, while satellite A follows a different path. During the optimisation, the RTN inter-satellite distance was evaluated to check for compliance with the collision threshold, as shown in Fig.~\ref{fig:EPM2SM_1_CA}. This shows how the evolution of the RTN distance between satellites A and B is equal to the A and C one, this is an effect of the symmetry of the optimal trajectories of satellites B and C. The control law required by the formation during the transition is shown in Fig.~\ref{fig:EPM2SM_1_thrust}. The resulting delta-v budget to perform the manoeuvre is about 6 cm/s for the three satellites for the Safe mode establishment.

\paragraph{CASE II:} This second case implements the radial-normal separation margin, introducing a phasing of the relative eccentricity/inclination vectors. The nominal formation aperture of 13 m was increased to 50 m, as in Case I. On the contrary, the aperture plane was slightly inclined on the TN plane, introducing a radial separation of $+ 5$ m and $- 5$ m with respect to the TN plane, for satellites B and C, respectively. For this reason, a higher delta-v is required to maintain satellites B and C out of the TN plane. The introduction of a radial separation provides a passive safety in both RTN and RN inter-satellite distance computation, as shown in Fig.~\ref{fig:EPM2SM_2}. On the other hand, it introduces the risk of formation evaporation in case of a malfunctioning of the propulsive system, due to the separation in the radial direction. The optimal trajectory for the transition from EPM to SM is shown in Fig.~\ref{fig:EPM2SM_2} and \ref{fig:EPM2SM_ROEs_2}. The trajectories for satellites B and C are designed to place the satellites at different components in the radial direction: satellite B gains a radial separation of $– 5$ m, while satellite C moves to $+ 5$ m in the radial direction. 
On the other hand, satellite A follows a different path and remains with a null component in the radial direction. During the optimisation, both the RTN and the RN inter-satellite distance was evaluated to check for compliance with the collision threshold, as shown in Fig.~\ref{fig:EPM2SM_2_CA}. Specifically, the RN distance was required to be compliant to the collision threshold only at the final time instant. The control law required by the formation during the transition is shown in Fig.~\ref{fig:EPM2SM_2_thrust}. Note that in this second case, the delta-v required for the manoeuvre, about 6.34 cm/s for A and 11.3 cm/s for B and C, is higher than Case I, due to the transversal component of the acceleration needed to move satellite B and C out of the TN plane.

\subsection{Thruster failure detection of one satellite in the nominal geometry formation}
\begin{figure}[!htb]
	\centering
	\begin{subfigure}[b]{0.3\textwidth}
		\centering
		\includegraphics[width=\textwidth]{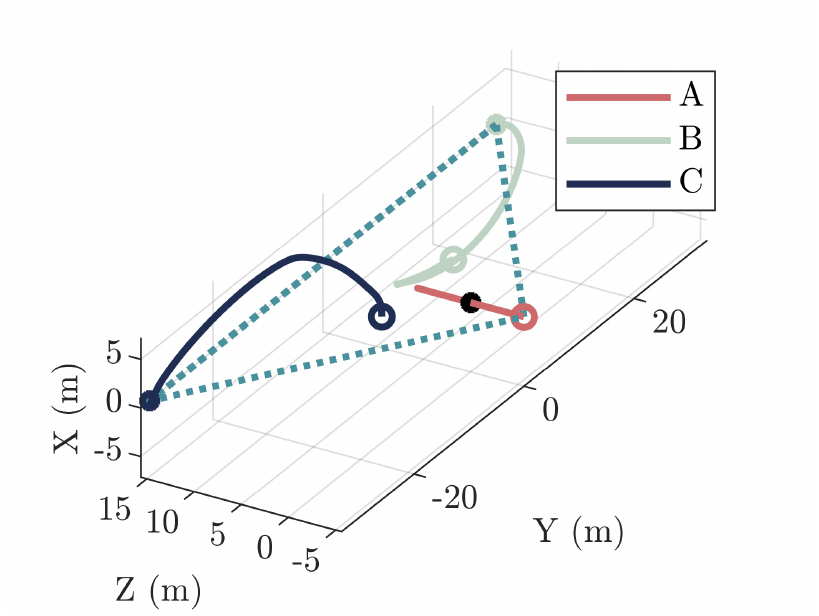}
		\caption{Optimal trajectory.}
		\label{fig:satA_fail}
	\end{subfigure}
	\hfill
	\begin{subfigure}[b]{0.3\textwidth}
		\centering
		\includegraphics[width=\textwidth]{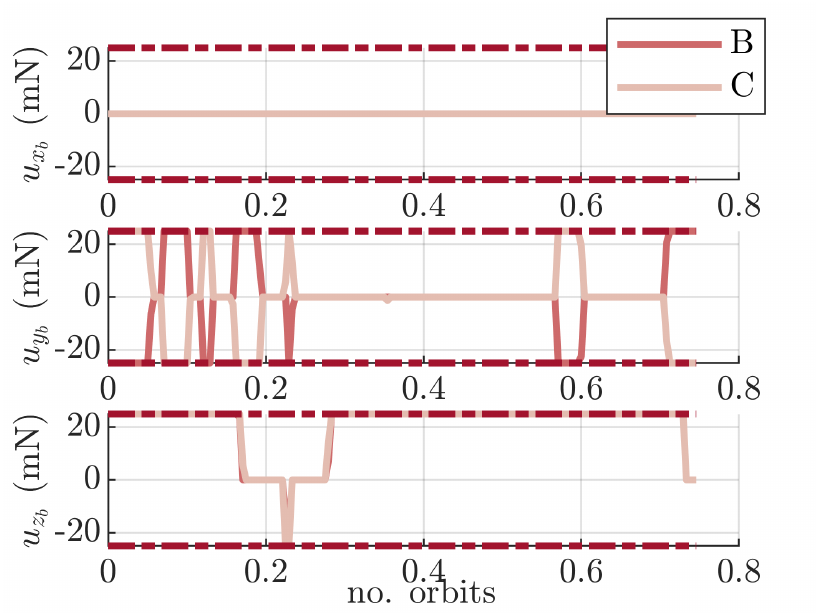}
		\caption{Control effort for satellite B and C.}
		\label{fig:satA_fail_thrust}
	\end{subfigure}	
	\hfill
	\begin{subfigure}[b]{0.3\textwidth}
		\centering
		\includegraphics[width=\textwidth]{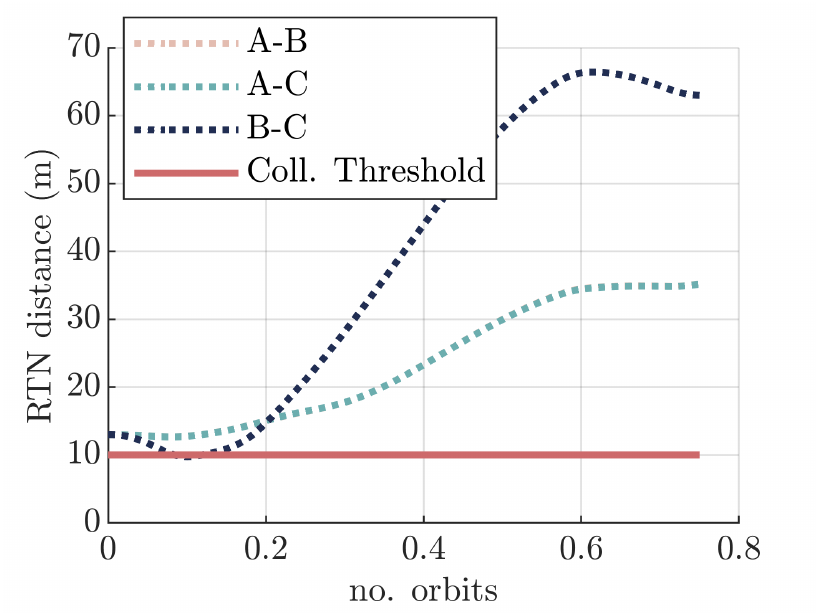}
		\caption{Inter-satellite distance during the manoeuvre.}
		\label{fig:satA_fail_CA}
	\end{subfigure}	
	\caption{Optimal manoeuvre for in case the propulsive system of the satellite A fails. The collision avoidance manoeuvre should be implemented by the satellite B and C as soon as the failure is detected.}
	\label{fig:satA_failure_varius}
\end{figure}

This section presents the manoeuvre to be implemented in case of a malfunctioning of the propulsion system of one satellite in the formation. The case of malfunctioning of satellite A is described. If the on-board low thrust control undergoes a non-nominal behaviour or a failure is detected, an inter-satellite collision avoidance manoeuvre shall be implemented. Specifically, the case under analysis requires an instantaneous manoeuvre for both satellites B and C in less than half of the orbital period. Therefore, when the failure of the propulsive system is detected in satellite A, satellites B and C should automatically implement the priority of actions to automatically transit to a safe region, to avoid any possible collision.
The first manoeuvre to be implemented should be fast enough to exit from the collision region. Then once the collision has been avoided, the need to pass to the two-satellite backup formation is evaluated, and, in that case, the consequent reconfiguration could be implemented. 
The final position of B and C, after the collision avoidance manoeuvre, is selected according to the safe mode formation geometries defined in Section \ref{sec:safe_mode}. In particular, the transition to Case I is considered. 
The control thrust needed by satellites B and C is shown in Fig.~\ref{fig:satA_fail_thrust}, respecting the need of using only the thrust in the normal and transversal directions. A real-time RTN and RN inter-satellite distance is evaluated during the manoeuvre, to ensure no collision in the formation. The collision threshold evaluation is reported in Fig.~\ref{fig:satA_fail_CA}. Once the first collision avoidance manoeuvre is implemented, the formation should run an analysis to understand and detect the causes of the failure on satellite A. At this point, if satellite A can recover from the failure, satellites B and C should manoeuvre to reconfigure the nominal Earth pointing formation.
On the contrary, if the failure of satellite A cannot be recovered, satellites B and C should manoeuvre to switch to the backup two-satellite formation.
The second possibility should implement a low thrust transfer to move in a safe position with respect to the failed satellite. The real mission scenario needs to implement a prediction of the future position of satellite A to define a safe area, where the backup two-satellite formation could continue the nominal operations. It should also investigate the possibility to perform a collision avoidance manoeuvre with satellite A after some propagation time.
A similar strategy is implemented for the cases when satellite B or C fails.


\section{Conclusions}

In this paper, we show a fast approach to perform the reconfiguration of multiple satellites formation in different operational scenarios.
The manoeuvring plan is defined via the optimisation of the control problem in the convex
form. This methodology allows a fast resolution of the system (2 sec to 5 sec for the cases presented),
by discretising the problem. Moreover, the methodology applied, ensures
the application of a delta-v only at optimal time instants, saving propellant consumption. Specifically, the reconfiguration problem is approached as a minimisation problem. The cost function aims at minimising the total delta-v for the manoeuvrers, while fulfilling the maximum thrust constraint. The explicit derivation of the constraint from the system dynamics, initial conditions, and final conditions support the software implementation of the proposed strategy. A convex approach is computationally less expensive than
an optimisation including the integration of the dynamics at each time step. This could also be
envisioned as a proper method to be implemented in the onboard software for formation manoeuvrer implementation. The performances of the convex approach are presented for a different number of satellites in the formation, up to a maximum of 12 spacecraft. This could provide a baseline for larger formations in terms of computational time.

The approach presents possible strategies for the reconfiguration manoeuvre in the typical operational scenarios of a remote sensing mission, as FFLAS. First, the implementation of the optimal transition between the earth pointing and the cold sky pointing mode is presented. The latter is essential for the calibration of the scientific instrument (the L-band interferometer), which is required at least once a month. Thus, it is important to propose optimal delta-v trajectories, to reduce the overall amount of propellant on-board. Moreover, due to the close formation geometries, with inter-satellite distance in the order of tens of metres, we presented a  fast and fuel-saving trajectory for the transition to a safe mode. This ensures the possibility to automatically deal with non-nominal situations in orbit and to minimise the collision risk among the satellites. 

The work presented was implemented at Politecnico di Milano, for the FFLAS mission concept, under the supervision of ESA ESTEC. Specifically, it was identified the need to design fuel-optimal manoeuvre for the nominal and non-nominal operation scenarios. During the analysis, the necessity to maintain a correct orientation among the satellites was considered, to guarantee the best performances both in terms of scientific payload and the inter-satellite telecommunication link.


\section*{Acknowledgements}
The work presented in this paper was co-founded by the European Space Agency (Contract No. 4000128576/19) and by the European Research Council (ERC) under the European Unions Horizon 2020 research and innovation program (grant agreement No. 679086 COMPASS). The view expressed in this paper can in no way be taken to reflect the official opinion of the European Space Agency. 
The contribution of Dr Gabriella Gaias is funded by the European Union’s Horizon 2020 research and innovation program under the Marie-Sklodowska Curie grant ReMoVE (grant agreement No. 793361).

The authors acknowledge the contribution of Dr Berthyl Duesmann and Dr Itziar Bara, orbital experts at ESA European Space Research and Technology Centre (ESTEC).
The authors also would like to acknowledge Dr Miguel Piera’s support, from Airbus Space Espa$\tilde{\textrm{n}}$a, leading the contract.

\appendix
\section{State transition matrix including Earth's oblateness}
\label{appendixA}
The relative mean motion is described through the model proposed by \cite{gaias2018semi}, which includes the primary perturbation in the LEO environment by the non-homogeneous Earth's mass distribution, in relative orbital elements representation. In this work, only the effect of the second-order term of the zonal harmonic ($J_2$) is considered. The state transition matrix that describes the linearized model for a generic eccentric reference orbit is given by: 
\begin{equation}
	\mathbf{A}_\alpha = \left[
	\begin{array}{cccccc}
	0	& 0 & 0 & 0 & 0 & 0 \\
	a\,G_{2,a}	& 0 & \cos \omega\,G_{2,e} & \sin \omega \,G_{2,e} & G_{2,i} & 0 \\
	a\,G_{3,a}	& 0 & \cos \omega\,G_{3,e} & \sin \omega \,G_{3,e} & G_{3,i} & 0 \\
	a\,G_{4,a}	& 0 & \cos \omega\,G_{4,e} & \sin \omega \,G_{4,e} & G_{4,i} & 0  \\
	0	& 0  & 0 & 0 & 0 & 0 \\
	a\,G_{6,a}	& 0 & \cos \omega\,G_{6,e} & \sin \omega \,G_{6,e} & G_{6,i} & 0 
	\end{array}	
	\right]
\end{equation}
Where the coefficients $G_{i,\alpha_j}$ is defined by the following notation:
\begin{equation}
	G_{i,\alpha_j} = \frac{\partial g_i}{\partial \alpha_j}
\end{equation}
Where only the partial with respect to $a$, $e$, $\omega$, and $i$ should be computed: 
\begin{equation}
	\left\{
	\begin{array}{l}
	g_2= \dot{M} + \dot{\omega}	 + \dot{\Omega}\cos i_c\\
	g_3 = - e\, \sin \omega	\, \dot{\omega}\\
	g_4 = + e\, \cos \omega	\, \dot{\omega}	\\
	g_6 = \dot{\Omega}\sin i_c	
	\end{array}
	\right.
\end{equation}
The corresponding value for the partial derivatives of $\dot{\Omega},\,\dot{\omega}$ or $\dot{M}$ are reported in \cite{gaias2018semi}.  These parameters account for the $J_2$ perturbing effect, which can be simplified considering a quasi-circular reference orbit.

\bibliographystyle{model5-names}
\biboptions{authoryear}
\bibliography{refs}

\begin{thebibliography}{40}
\expandafter\ifx\csname natexlab\endcsname\relax\def\natexlab#1{#1}\fi
\providecommand{\url}[1]{\texttt{#1}}
\providecommand{\href}[2]{#2}
\providecommand{\path}[1]{#1}
\providecommand{\DOIprefix}{doi:}
\providecommand{\ArXivprefix}{arXiv:}
\providecommand{\URLprefix}{URL: }
\providecommand{\Pubmedprefix}{pmid:}
\providecommand{\doi}[1]{\href{http://dx.doi.org/#1}{\path{#1}}}
\providecommand{\Pubmed}[1]{\href{pmid:#1}{\path{#1}}}
\providecommand{\bibinfo}[2]{#2}
\ifx\xfnm\relax \def\xfnm[#1]{\unskip,\space#1}\fi
\bibitem[{Acikmese et~al.(2006)Acikmese, Scharf, Hadaegh \&
  Murray}]{acikmese2006convex}
\bibinfo{author}{Acikmese, B.}, \bibinfo{author}{Scharf, D.},
  \bibinfo{author}{Hadaegh, F.}, \& \bibinfo{author}{Murray, E.}
  (\bibinfo{year}{2006}).
\newblock \bibinfo{title}{A convex guidance algorithm for formation
  reconfiguration}.
\newblock In {\it \bibinfo{booktitle}{AIAA Guidance, Navigation, and Control
  Conference and Exhibit}\/} (p. \bibinfo{pages}{6070}).
\newblock \DOIprefix\doi{10.2514/6.2006-6070}.
\bibitem[{Alfriend et~al.(2009)Alfriend, Vadali, Gurfil, How \&
  Breger}]{alfriend2009spacecraft}
\bibinfo{author}{Alfriend, K.}, \bibinfo{author}{Vadali, S.~R.},
  \bibinfo{author}{Gurfil, P.}, \bibinfo{author}{How, J.}, \&
  \bibinfo{author}{Breger, L.} (\bibinfo{year}{2009}).
\newblock {\it \bibinfo{title}{Spacecraft formation flying: Dynamics, control
  and navigation}\/} volume~\bibinfo{volume}{2}.
\newblock \bibinfo{publisher}{Elsevier}.
\bibitem[{Armellin et~al.(2004)Armellin, Massari \&
  Finzi}]{armellin2004optimal}
\bibinfo{author}{Armellin, R.}, \bibinfo{author}{Massari, M.}, \&
  \bibinfo{author}{Finzi, A.~E.} (\bibinfo{year}{2004}).
\newblock \bibinfo{title}{Optimal formation flying reconfiguration and station
  keeping maneuvers using low thrust propulsion}.
\newblock In {\it \bibinfo{booktitle}{Proceedings of the 18th International
  Symposium on Space Flight Dynamics (ESA SP-548)}\/} \bibinfo{number}{548}
  (pp. \bibinfo{pages}{429--434}).
\bibitem[{Bandyopadhyay et~al.(2016)Bandyopadhyay, Foust, Subramanian, Chung \&
  Hadaegh}]{bandyopadhyay2016review}
\bibinfo{author}{Bandyopadhyay, S.}, \bibinfo{author}{Foust, R.},
  \bibinfo{author}{Subramanian, G.~P.}, \bibinfo{author}{Chung, S.-J.}, \&
  \bibinfo{author}{Hadaegh, F.~Y.} (\bibinfo{year}{2016}).
\newblock \bibinfo{title}{Review of formation flying and constellation missions
  using nanosatellites}.
\newblock {\it \bibinfo{journal}{Journal of Spacecraft and Rockets}\/},  {\it
  \bibinfo{volume}{53}\/}\bibinfo{issue}{(3)}, \bibinfo{pages}{567--578}.
  \DOIprefix\doi{10.2514/1.A33291}.
\bibitem[{Boyd et~al.(2004)Boyd, Boyd \& Vandenberghe}]{boyd2004convex}
\bibinfo{author}{Boyd, S.}, \bibinfo{author}{Boyd, S.~P.}, \&
  \bibinfo{author}{Vandenberghe, L.} (\bibinfo{year}{2004}).
\newblock {\it \bibinfo{title}{Convex optimization}\/}.
\newblock \bibinfo{publisher}{Cambridge university press}.
\bibitem[{Clohessy \& Wiltshire(1960)}]{clohessy1960terminal}
\bibinfo{author}{Clohessy, W.}, \& \bibinfo{author}{Wiltshire, R.}
  (\bibinfo{year}{1960}).
\newblock \bibinfo{title}{Terminal guidance system for satellite rendezvous}.
\newblock {\it \bibinfo{journal}{Journal of the Aerospace Sciences}\/},  {\it
  \bibinfo{volume}{27}\/}\bibinfo{issue}{(9)}, \bibinfo{pages}{653--658}.
  \DOIprefix\doi{10.2514/8.8704}.
\bibitem[{D’Amico(2005)}]{d2005relative}
\bibinfo{author}{D’Amico, S.} (\bibinfo{year}{2005}).
\newblock \bibinfo{title}{Relative orbital elements as integration constants of
  hill’s equations}.
\newblock {\it \bibinfo{journal}{DLR, TN}\/},  (pp. \bibinfo{pages}{05--08}).
\bibitem[{Gaias \& Colombo(2018)}]{gaias2018semi}
\bibinfo{author}{Gaias, G.}, \& \bibinfo{author}{Colombo, C.}
  (\bibinfo{year}{2018}).
\newblock \bibinfo{title}{Semi-analytical framework for precise relative motion
  in low earth orbits}.
\newblock In {\it \bibinfo{booktitle}{7th International Conference on
  Astrodynamics Tools and Techniques (ICATT)}\/} (pp. \bibinfo{pages}{1--10}).
\bibitem[{Gaias et~al.(2020)Gaias, Colombo \& Lara}]{gaias2020analytical}
\bibinfo{author}{Gaias, G.}, \bibinfo{author}{Colombo, C.}, \&
  \bibinfo{author}{Lara, M.} (\bibinfo{year}{2020}).
\newblock \bibinfo{title}{Analytical framework for precise relative motion in
  low earth orbits}.
\newblock {\it \bibinfo{journal}{Journal of Guidance, Control, and
  Dynamics}\/},  {\it \bibinfo{volume}{43}\/}\bibinfo{issue}{(5)},
  \bibinfo{pages}{915--927}. \DOIprefix\doi{10.2514/1.G004716}.
\bibitem[{Gaias \& Lovera(2020)}]{gaias2020safe}
\bibinfo{author}{Gaias, G.}, \& \bibinfo{author}{Lovera, M.}
  (\bibinfo{year}{2020}).
\newblock \bibinfo{title}{Safe trajectory design for close proximity
  operations}.
\newblock In {\it \bibinfo{booktitle}{2020 AAS/AIAA Astrodynamics Specialist
  Conference}\/} (pp. \bibinfo{pages}{1--15}).
\bibitem[{Grant \& Boyd(2008)}]{grant2008graph}
\bibinfo{author}{Grant, M.}, \& \bibinfo{author}{Boyd, S.}
  (\bibinfo{year}{2008}).
\newblock \bibinfo{title}{Graph implementations for nonsmooth convex programs}.
\newblock In \bibinfo{editor}{V.~Blondel}, \bibinfo{editor}{S.~Boyd}, \&
  \bibinfo{editor}{H.~Kimura} (Eds.), {\it \bibinfo{booktitle}{Recent Advances
  in Learning and Control}\/} Lecture Notes in Control and Information Sciences
  (pp. \bibinfo{pages}{95--110}).
\newblock \bibinfo{publisher}{Springer-Verlag Limited}.
\newblock \DOIprefix\doi{10.1007/978-1-84800-155-8_7}.
\bibitem[{Grant et~al.(2013)Grant, Boyd \& Ye}]{grant2013cvx}
\bibinfo{author}{Grant, M.}, \bibinfo{author}{Boyd, S.}, \&
  \bibinfo{author}{Ye, Y.} (\bibinfo{year}{2013}).
\newblock \bibinfo{title}{Cvx: Matlab software for disciplined convex
  programming, version 2.0 beta}.
\newblock \bibinfo{howpublished}{\url{http://cvxr.com/cvx}}.
\bibitem[{{Gurobi Optimization, LLC}(2021)}]{gurobi}
\bibinfo{author}{{Gurobi Optimization, LLC}} (\bibinfo{year}{2021}).
\newblock \bibinfo{title}{{Gurobi Optimizer Reference Manual}}.
\newblock \URLprefix \url{https://www.gurobi.com}.
\bibitem[{Hill(1878)}]{hill1878researches}
\bibinfo{author}{Hill, G.~W.} (\bibinfo{year}{1878}).
\newblock \bibinfo{title}{Researches in the lunar theory}.
\newblock {\it \bibinfo{journal}{American journal of Mathematics}\/},  {\it
  \bibinfo{volume}{1}\/}\bibinfo{issue}{(1)}, \bibinfo{pages}{5--26}.
  \DOIprefix\doi{10.2307/2369430}.
\bibitem[{Izzo et~al.(2003)Izzo, Sabatini \& Valente}]{izzo2003new}
\bibinfo{author}{Izzo, D.}, \bibinfo{author}{Sabatini, M.}, \&
  \bibinfo{author}{Valente, C.} (\bibinfo{year}{2003}).
\newblock \bibinfo{title}{A new linear model describing formation flying
  dynamics under j2 effects}.
\newblock In {\it \bibinfo{booktitle}{Proceedings of the 17th AIDAA National
  Congress}\/} (pp. \bibinfo{pages}{15--19}).
\newblock volume~\bibinfo{volume}{1}.
\bibitem[{Kerr et~al.(2016)Kerr, Al-Yaari, Rodriguez-Fernandez, Parrens,
  Molero, Leroux, Bircher, Mahmoodi, Mialon, Richaume
  et~al.}]{kerr2016overview}
\bibinfo{author}{Kerr, Y.~H.}, \bibinfo{author}{Al-Yaari, A.},
  \bibinfo{author}{Rodriguez-Fernandez, N.}, \bibinfo{author}{Parrens, M.},
  \bibinfo{author}{Molero, B.}, \bibinfo{author}{Leroux, D.},
  \bibinfo{author}{Bircher, S.}, \bibinfo{author}{Mahmoodi, A.},
  \bibinfo{author}{Mialon, A.}, \bibinfo{author}{Richaume, P.} et~al.
  (\bibinfo{year}{2016}).
\newblock \bibinfo{title}{Overview of smos performance in terms of global soil
  moisture monitoring after six years in operation}.
\newblock {\it \bibinfo{journal}{Remote Sensing of Environment}\/},  {\it
  \bibinfo{volume}{180}\/}, \bibinfo{pages}{40--63}.
  \DOIprefix\doi{10.1016/j.rse.2016.02.042}.
\bibitem[{{Kerr} et~al.(2020){Kerr}, {Rodriguez-Fernandez}, {Anterrieu},
  {Escorihuela}, {Drusch}, {Closa}, {Zurita}, {Cabot}, {Amiot}, {Bindlish} \&
  {O'Neill}}]{kerrLband}
\bibinfo{author}{{Kerr}, Y.~H.}, \bibinfo{author}{{Rodriguez-Fernandez}, N.},
  \bibinfo{author}{{Anterrieu}, E.}, \bibinfo{author}{{Escorihuela}, M.~J.},
  \bibinfo{author}{{Drusch}, M.}, \bibinfo{author}{{Closa}, J.},
  \bibinfo{author}{{Zurita}, A.}, \bibinfo{author}{{Cabot}, F.},
  \bibinfo{author}{{Amiot}, T.}, \bibinfo{author}{{Bindlish}, R.}, \&
  \bibinfo{author}{{O'Neill}, P.} (\bibinfo{year}{2020}).
\newblock \bibinfo{title}{The next generation of l band radiometry: User's
  requirements and technical solutions}.
\newblock In {\it \bibinfo{booktitle}{IGARSS 2020 - 2020 IEEE International
  Geoscience and Remote Sensing Symposium}\/} (pp.
  \bibinfo{pages}{5974--5977}).
\newblock \DOIprefix\doi{10.1109/IGARSS39084.2020.9324452}.
\bibitem[{Koenig \& D'Amico(2020)}]{koenig2020fast}
\bibinfo{author}{Koenig, A.}, \& \bibinfo{author}{D'Amico, S.}
  (\bibinfo{year}{2020}).
\newblock \bibinfo{title}{Fast algorithm for fuel-optimal impulsive control of
  linear systems with time-varying cost}.
\newblock {\it \bibinfo{journal}{IEEE Transactions on Automatic Control}\/},
  (pp. \bibinfo{pages}{1--1}). \DOIprefix\doi{10.1109/TAC.2020.3027804}.
\bibitem[{Koenig \& D'Amico(2018)}]{koenig2018safe}
\bibinfo{author}{Koenig, A.~W.}, \& \bibinfo{author}{D'Amico, S.}
  (\bibinfo{year}{2018}).
\newblock \bibinfo{title}{Safe spacecraft swarm deployment and acquisition in
  perturbed near-circular orbits subject to operational constraints}.
\newblock {\it \bibinfo{journal}{Acta Astronautica}\/},  {\it
  \bibinfo{volume}{153}\/}, \bibinfo{pages}{297--310}.
  \DOIprefix\doi{10.1016/j.actaastro.2018.01.037}.
\bibitem[{Krieger et~al.(2007)Krieger, Moreira, Fiedler, Hajnsek, Werner,
  Younis \& Zink}]{krieger2007tandem}
\bibinfo{author}{Krieger, G.}, \bibinfo{author}{Moreira, A.},
  \bibinfo{author}{Fiedler, H.}, \bibinfo{author}{Hajnsek, I.},
  \bibinfo{author}{Werner, M.}, \bibinfo{author}{Younis, M.}, \&
  \bibinfo{author}{Zink, M.} (\bibinfo{year}{2007}).
\newblock \bibinfo{title}{Tandem-x: A satellite formation for high-resolution
  sar interferometry}.
\newblock {\it \bibinfo{journal}{IEEE Transactions on Geoscience and Remote
  Sensing}\/},  {\it \bibinfo{volume}{45}\/}\bibinfo{issue}{(11)},
  \bibinfo{pages}{3317--3341}. \DOIprefix\doi{10.1109/TGRS.2007.900693}.
\bibitem[{Leitner(2004)}]{leitner2004formation}
\bibinfo{author}{Leitner, J.} (\bibinfo{year}{2004}).
\newblock \bibinfo{title}{Formation flying-the future of remote sensing from
  space}.
\newblock In {\it \bibinfo{booktitle}{18th International Symposium on Space
  Flight Dynamics}\/} (p. \bibinfo{pages}{621}).
\newblock volume \bibinfo{volume}{548}.
\bibitem[{{Martín-Neira} et~al.(2020){Martín-Neira}, {Suess}, {Karafolas},
  {Piironen}, {Deborgies}, {Catalán}, {Vilaseca}, {Montero}, {Puertolas},
  {Outumuro}, {Corbella}, {Durán}, {Duffo}, {Materni}, {Mengual}, {Piqueras},
  {Olea}, {Solana}, {Closa}, {Zurita}, {Ramírez}, {Breinbjerg}, {Bjørstorp},
  {Kaslis}, {Kristensen}, {Oliva}, {Onrubia}, {Camps} \& {Querol}}]{Neira2020}
\bibinfo{author}{{Martín-Neira}, M.}, \bibinfo{author}{{Suess}, M.},
  \bibinfo{author}{{Karafolas}, N.}, \bibinfo{author}{{Piironen}, P.},
  \bibinfo{author}{{Deborgies}, F.}, \bibinfo{author}{{Catalán}, A.},
  \bibinfo{author}{{Vilaseca}, R.}, \bibinfo{author}{{Montero}, J.},
  \bibinfo{author}{{Puertolas}, M.}, \bibinfo{author}{{Outumuro}, D.},
  \bibinfo{author}{{Corbella}, I.}, \bibinfo{author}{{Durán}, I.},
  \bibinfo{author}{{Duffo}, N.}, \bibinfo{author}{{Materni}, R.},
  \bibinfo{author}{{Mengual}, T.}, \bibinfo{author}{{Piqueras}, M.~A.},
  \bibinfo{author}{{Olea}, A.}, \bibinfo{author}{{Solana}, A.},
  \bibinfo{author}{{Closa}, J.}, \bibinfo{author}{{Zurita}, A.},
  \bibinfo{author}{{Ramírez}, J.~I.}, \bibinfo{author}{{Breinbjerg}, O.},
  \bibinfo{author}{{Bjørstorp}, J.~M.}, \bibinfo{author}{{Kaslis}, K.},
  \bibinfo{author}{{Kristensen}, S.~S.}, \bibinfo{author}{{Oliva}, R.},
  \bibinfo{author}{{Onrubia}, R.}, \bibinfo{author}{{Camps}, A.}, \&
  \bibinfo{author}{{Querol}, J.} (\bibinfo{year}{2020}).
\newblock \bibinfo{title}{Technology developments for an advanced l-band
  radiometer mission}.
\newblock In {\it \bibinfo{booktitle}{IGARSS 2020 - 2020 IEEE International
  Geoscience and Remote Sensing Symposium}\/} (pp.
  \bibinfo{pages}{6507--6510}).
\newblock \DOIprefix\doi{10.1109/IGARSS39084.2020.9324378}.
\bibitem[{Mitani \& Yamakawa(2013)}]{mitani2013continuous}
\bibinfo{author}{Mitani, S.}, \& \bibinfo{author}{Yamakawa, H.}
  (\bibinfo{year}{2013}).
\newblock \bibinfo{title}{Continuous-thrust transfer with control magnitude and
  direction constraints using smoothing techniques}.
\newblock {\it \bibinfo{journal}{Journal of guidance, control, and
  dynamics}\/},  {\it \bibinfo{volume}{36}\/}\bibinfo{issue}{(1)},
  \bibinfo{pages}{163--174}. \DOIprefix\doi{10.2514/1.56882}.
\bibitem[{Montenbruck et~al.(2015)Montenbruck, Schmid, Mercier, Steigenberger,
  Noll, Fatkulin, Kogure \& Ganeshan}]{montenbruck2015gnss}
\bibinfo{author}{Montenbruck, O.}, \bibinfo{author}{Schmid, R.},
  \bibinfo{author}{Mercier, F.}, \bibinfo{author}{Steigenberger, P.},
  \bibinfo{author}{Noll, C.}, \bibinfo{author}{Fatkulin, R.},
  \bibinfo{author}{Kogure, S.}, \& \bibinfo{author}{Ganeshan, A.~S.}
  (\bibinfo{year}{2015}).
\newblock \bibinfo{title}{Gnss satellite geometry and attitude models}.
\newblock {\it \bibinfo{journal}{Advances in Space Research}\/},  {\it
  \bibinfo{volume}{56}\/}\bibinfo{issue}{(6)}, \bibinfo{pages}{1015--1029}.
  \DOIprefix\doi{10.1016/j.asr.2015.06.019}.
\bibitem[{Moreira et~al.(2015)Moreira, Krieger, Hajnsek, Papathanassiou,
  Younis, Lopez-Dekker, Huber, Villano, Pardini, Eineder
  et~al.}]{moreira2015tandem}
\bibinfo{author}{Moreira, A.}, \bibinfo{author}{Krieger, G.},
  \bibinfo{author}{Hajnsek, I.}, \bibinfo{author}{Papathanassiou, K.},
  \bibinfo{author}{Younis, M.}, \bibinfo{author}{Lopez-Dekker, P.},
  \bibinfo{author}{Huber, S.}, \bibinfo{author}{Villano, M.},
  \bibinfo{author}{Pardini, M.}, \bibinfo{author}{Eineder, M.} et~al.
  (\bibinfo{year}{2015}).
\newblock \bibinfo{title}{Tandem-l: A highly innovative bistatic sar mission
  for global observation of dynamic processes on the earth's surface}.
\newblock {\it \bibinfo{journal}{IEEE Geoscience and Remote Sensing
  Magazine}\/},  {\it \bibinfo{volume}{3}\/}\bibinfo{issue}{(2)},
  \bibinfo{pages}{8--23}. \DOIprefix\doi{10.1109/MGRS.2015.2437353}.
\bibitem[{Morgan et~al.(2014)Morgan, Chung \& Hadaegh}]{Morgan2014mpc}
\bibinfo{author}{Morgan, D.}, \bibinfo{author}{Chung, S.-J.}, \&
  \bibinfo{author}{Hadaegh, F.~Y.} (\bibinfo{year}{2014}).
\newblock \bibinfo{title}{Model predictive control of swarms of spacecraft
  using sequential convex programming}.
\newblock {\it \bibinfo{journal}{Journal of Guidance, Control, and
  Dynamics}\/},  {\it \bibinfo{volume}{37}\/}\bibinfo{issue}{(6)},
  \bibinfo{pages}{1725--1740}. \DOIprefix\doi{10.2514/1.G000218}.
\bibitem[{Randall et~al.(2017)Randall, Lewis \& Clark}]{randall2017qinetiq}
\bibinfo{author}{Randall, P.~N.}, \bibinfo{author}{Lewis, R.~A.}, \&
  \bibinfo{author}{Clark, S.~D.} (\bibinfo{year}{2017}).
\newblock \bibinfo{title}{Qinetiq t5 based electric propulsion system and
  architectural options for future applications}.
\newblock In {\it \bibinfo{booktitle}{35th International Electric Propulsion
  Conference}\/}.
\bibitem[{Rowell(2002)}]{rowell2002state}
\bibinfo{author}{Rowell, D.} (\bibinfo{year}{2002}).
\newblock \bibinfo{title}{State-space representation of lti systems}, .
\newblock \URLprefix \url{http://web.mit.edu/2.14/www/Handouts/StateSpace.pdf}.
\bibitem[{{Sabatini} \& {Palmerini}(2008)}]{Sabatini2008}
\bibinfo{author}{{Sabatini}, M.}, \& \bibinfo{author}{{Palmerini}, G.~B.}
  (\bibinfo{year}{2008}).
\newblock \bibinfo{title}{Linearized formation-flying dynamics in a perturbed
  orbital environment}.
\newblock In {\it \bibinfo{booktitle}{2008 IEEE Aerospace Conference}\/} (pp.
  \bibinfo{pages}{1--13}).
\newblock \DOIprefix\doi{10.1109/AERO.2008.4526271}.
\bibitem[{Sabatini \& Palmerini(2009)}]{sabatini2009collective}
\bibinfo{author}{Sabatini, M.}, \& \bibinfo{author}{Palmerini, G.~B.}
  (\bibinfo{year}{2009}).
\newblock \bibinfo{title}{Collective control of spacecraft swarms for space
  exploration}.
\newblock {\it \bibinfo{journal}{Celestial Mechanics and Dynamical
  Astronomy}\/},  {\it \bibinfo{volume}{105}\/}\bibinfo{issue}{(1)},
  \bibinfo{pages}{229--244}. \DOIprefix\doi{10.1007/s10569-009-9183-8}.
\bibitem[{Sarno et~al.(2020)Sarno, Guo, D'Errico \& Gill}]{sarno2020guidance}
\bibinfo{author}{Sarno, S.}, \bibinfo{author}{Guo, J.},
  \bibinfo{author}{D'Errico, M.}, \& \bibinfo{author}{Gill, E.}
  (\bibinfo{year}{2020}).
\newblock \bibinfo{title}{A guidance approach to satellite formation
  reconfiguration based on convex optimization and genetic algorithms}.
\newblock {\it \bibinfo{journal}{Advances in Space Research}\/},  {\it
  \bibinfo{volume}{65}\/}\bibinfo{issue}{(8)}, \bibinfo{pages}{2003--2017}.
  \DOIprefix\doi{10.1016/j.asr.2020.01.033}.
\bibitem[{Scala et~al.(2020{\natexlab{a}})Scala, Gaias, Colombo, Martin~Neira
  et~al.}]{scala2020three}
\bibinfo{author}{Scala, F.}, \bibinfo{author}{Gaias, G.},
  \bibinfo{author}{Colombo, C.}, \bibinfo{author}{Martin~Neira, M.} et~al.
  (\bibinfo{year}{2020}{\natexlab{a}}).
\newblock \bibinfo{title}{Three satellites formation flying: Deployment and
  formation acquisition using relative orbital elements}.
\newblock In {\it \bibinfo{booktitle}{2020 AAS/AIAA Astrodynamics Specialist
  Conference}\/} (pp. \bibinfo{pages}{1--17}).
\bibitem[{Scala et~al.(2020{\natexlab{b}})Scala, Gaias, Colombo \&
  Martìn-Neira}]{scala2020formation}
\bibinfo{author}{Scala, F.}, \bibinfo{author}{Gaias, G.},
  \bibinfo{author}{Colombo, C.}, \& \bibinfo{author}{Martìn-Neira, M.}
  (\bibinfo{year}{2020}{\natexlab{b}}).
\newblock \bibinfo{title}{Formation flying l-band aperture synthesis: Design
  challenges and innovative formation architecture concept}.
\newblock {\it \bibinfo{journal}{Proceedings of the International Astronautical
  Congress, IAC}\/},  {\it \bibinfo{volume}{2020-October}\/}.
\bibitem[{Schweighart \& Sedwick(2002)}]{schweighart2002high}
\bibinfo{author}{Schweighart, S.~A.}, \& \bibinfo{author}{Sedwick, R.~J.}
  (\bibinfo{year}{2002}).
\newblock \bibinfo{title}{High-fidelity linearized j model for satellite
  formation flight}.
\newblock {\it \bibinfo{journal}{Journal of Guidance, Control, and
  Dynamics}\/},  {\it \bibinfo{volume}{25}\/}\bibinfo{issue}{(6)},
  \bibinfo{pages}{1073--1080}. \DOIprefix\doi{10.2514/2.4986}.
\bibitem[{Sturm(1999)}]{sturm1999using}
\bibinfo{author}{Sturm, J.~F.} (\bibinfo{year}{1999}).
\newblock \bibinfo{title}{Using sedumi 1.02, a matlab toolbox for optimization
  over symmetric cones}.
\newblock {\it \bibinfo{journal}{Optimization methods and software}\/},  {\it
  \bibinfo{volume}{11}\/}\bibinfo{issue}{(1-4)}, \bibinfo{pages}{625--653}.
  \DOIprefix\doi{10.1080/10556789908805766}.
\bibitem[{Tillerson et~al.(2002)Tillerson, Inalhan \& How}]{tillerson2002co}
\bibinfo{author}{Tillerson, M.}, \bibinfo{author}{Inalhan, G.}, \&
  \bibinfo{author}{How, J.~P.} (\bibinfo{year}{2002}).
\newblock \bibinfo{title}{Co-ordination and control of distributed spacecraft
  systems using convex optimization techniques}.
\newblock {\it \bibinfo{journal}{International Journal of Robust and Nonlinear
  Control: IFAC-Affiliated Journal}\/},  {\it
  \bibinfo{volume}{12}\/}\bibinfo{issue}{(2-3)}, \bibinfo{pages}{207--242}.
  \DOIprefix\doi{10.1002/rnc.683}.
\bibitem[{Toh et~al.(1999)Toh, Todd \& T{\"u}t{\"u}nc{\"u}}]{toh1999sdpt3}
\bibinfo{author}{Toh, K.-C.}, \bibinfo{author}{Todd, M.~J.}, \&
  \bibinfo{author}{T{\"u}t{\"u}nc{\"u}, R.~H.} (\bibinfo{year}{1999}).
\newblock \bibinfo{title}{Sdpt3—a matlab software package for semidefinite
  programming, version 1.3}.
\newblock {\it \bibinfo{journal}{Optimization methods and software}\/},  {\it
  \bibinfo{volume}{11}\/}\bibinfo{issue}{(1-4)}, \bibinfo{pages}{545--581}.
  \DOIprefix\doi{10.1080/10556789908805762}.
\bibitem[{Vadali et~al.(2000)Vadali, Alfriend \& Vaddi}]{vadali2000hill}
\bibinfo{author}{Vadali, S.}, \bibinfo{author}{Alfriend, K.~T.}, \&
  \bibinfo{author}{Vaddi, S.} (\bibinfo{year}{2000}).
\newblock \bibinfo{title}{Hill's equations, mean orbital elements, and
  formation flying of satellites}.
\newblock In {\it \bibinfo{booktitle}{The Richard H. Battin Astrodynamics
  Symposium, College Station, TX}\/} (pp. \bibinfo{pages}{187--203}).
\newblock volume \bibinfo{volume}{AAS 00-258}.
\bibitem[{Wermuth et~al.(2015)Wermuth, Gaias \& D’Amico}]{Wermuth2015}
\bibinfo{author}{Wermuth, M.}, \bibinfo{author}{Gaias, G.}, \&
  \bibinfo{author}{D’Amico, S.} (\bibinfo{year}{2015}).
\newblock \bibinfo{title}{Safe picosatellite release from a small satellite
  carrier}.
\newblock {\it \bibinfo{journal}{Journal of Spacecraft and Rockets}\/},  {\it
  \bibinfo{volume}{52}\/}\bibinfo{issue}{(5)}, \bibinfo{pages}{1338--1347}.
  \DOIprefix\doi{10.2514/1.A33036}.
\bibitem[{Zurita et~al.(2013)Zurita, Corbella, Mart{\'\i}n-Neira, Plaza, Torres
  \& Benito}]{Zurita2013}
\bibinfo{author}{Zurita, A.~M.}, \bibinfo{author}{Corbella, I.},
  \bibinfo{author}{Mart{\'\i}n-Neira, M.}, \bibinfo{author}{Plaza, M.~A.},
  \bibinfo{author}{Torres, F.}, \& \bibinfo{author}{Benito, F.~J.}
  (\bibinfo{year}{2013}).
\newblock \bibinfo{title}{Towards a smos operational mission:
  Smosops-hexagonal}.
\newblock {\it \bibinfo{journal}{IEEE Journal of Selected Topics in Applied
  Earth Observations and Remote Sensing}\/},  {\it
  \bibinfo{volume}{6}\/}\bibinfo{issue}{(3)}, \bibinfo{pages}{1769--1780}.
  \DOIprefix\doi{10.1109/JSTARS.2013.2265600}.

\end{thebibliography}

\end{document}